\documentclass[english]{article}
\usepackage{graphicx}
\usepackage[T1]{fontenc}
\usepackage[latin1]{inputenc}
\usepackage{geometry}
\usepackage{amsmath}
\usepackage{amssymb}

\makeatletter

\providecommand{\LyX}{L\kern-.1667em\lower.25em\hbox{Y}\kern-.125emX\@}

\newcounter{draft}
\usepackage{amssymb} 
\usepackage{ifthen}
\ifthenelse{\thedraft=1}{\usepackage[notref,notcite]{showkeys}}{}

\usepackage{hyperref}
\usepackage{cite} 

\numberwithin{equation}{section}

\newcommand{\eref}{\def\theequation{{\bf \thesection}.\arabic{equation}}}

\newcommand{\rem}[1]{\ifthenelse{\thedraft=1}{\textsc{\texttt{<< #1 >>}}}{}}

\newcommand{\bmul}{\begin{multicols}{2}[\vspace{-1cm}]}
\newcommand{\emul}{\end{multicols}\vspace{-0.5cm}}

\date{\rem{\today}}
\newcommand{\Teil}[2]{#2} 

\usepackage{babel}
\makeatother
\begin{document}
\Teil{Macros}{
\newcommand{\bs}[1]{\boldsymbol{#1}}

\newcommand{\mf}[1]{\mathfrak{#1} }

\newcommand{\mc}[1]{\mathcal{#1}}

\newcommand{\norm}[1]{{\parallel#1 \parallel}}

\newcommand{\Norm}[1]{\left\Vert #1 \right\Vert }

\newcommand{\partiell}[2]{\frac{\partial#1 }{\partial#2 }}

\newcommand{\Partiell}[2]{\left( \frac{\partial#1 }{\partial#2 }\right) }

\newcommand{\ola}[1]{\overleftarrow{#1}}

\newcommand{\lpartial}{\overleftarrow{\partial}}

\newcommand{\partl}[1]{\frac{\partial}{\partial#1}}

\newcommand{\partr}[1]{\frac{\lpartial}{\partial#1}}

\newcommand{\funktional}[2]{\frac{\delta#1 }{\delta#2 }}

\newcommand{\funktl}[1]{\frac{\delta}{\delta#1}}

\newcommand{\funktr}[1]{\frac{\ola{\delta}}{\delta#1}}

\newcommand{\de}{{\bf d}\!}

\newcommand{\es}{{\bf s}\!}

\newcommand{\dew}{\bs{d}^{\textrm{w}}\!}

\newcommand{\Lie}{\bs{\mc{L}}}

\newcommand{\Dorf}{\bs{\mc{D}}}

\newcommand{\pe}{\bs{\partial}}

\newcommand{\De}{\textrm{D}\!}

\newcommand{\total}[2]{\frac{\de#1 }{\de#2 }}

\newcommand{\Frac}[2]{\left( \frac{#1 }{#2 }\right) }

\newcommand{\To}{\rightarrow}
 
\newcommand{\ket}[1]{|#1 >}

\newcommand{\bra}[1]{<#1 |}

\newcommand{\Ket}[1]{\left| #1 \right\rangle }

\newcommand{\Bra}[1]{\left\langle #1 \right| }
 
\newcommand{\braket}[2]{<#1 |#2 >}

\newcommand{\Braket}[2]{\Bra{#1 }\left. #2 \right\rangle }

\newcommand{\kom}[2]{[#1 ,#2 ]}

\newcommand{\Kom}[2]{\left[ #1 ,#2 \right] }

\newcommand{\abs}[1]{\mid#1 \mid}

\newcommand{\Abs}[1]{\left| #1 \right| }

\newcommand{\erw}[1]{\langle#1\rangle}

\newcommand{\Erw}[1]{\left\langle #1 \right\rangle }

\newcommand{\bei}[2]{\left. #1 \right| _{#2 }}

\newcommand{\dann}{\Rightarrow}

\newcommand{\q}[1]{\underline{#1 }}

\newcommand{\hoch}[1]{{}^{#1 }}

\newcommand{\tief}[1]{{}_{#1 }}

\newcommand{\lqn}[1]{\lefteqn{#1}}

\newcommand{\os}[2]{\overset{\lqn{#1}}{#2}}

\newcommand{\us}[2]{\underset{\lqn{{\scriptstyle #2}}}{#1}}

\newcommand{\ous}[3]{\underset{#3}{\os{#1}{#2}}}

\newcommand{\zwek}[2]{\begin{array}{c}
#1\\
#2\end{array}}

\newcommand{\drek}[3]{\begin{array}{c}
#1\\
#2\\
#3\end{array}}

\newcommand{\UB}[2]{\underbrace{#1 }_{\le{#2 }}}

\newcommand{\OB}[2]{\overbrace{#1 }^{\le{#2 }}}

\newcommand{\tr}{\textrm{tr}\,}

\newcommand{\Tr}{\textrm{Tr}\,}

\newcommand{\Det}{\textrm{Det}\,}

\newcommand{\diag}{\textrm{diag}\,}

\newcommand{\Diag}{\textrm{Diag}\,}

\newcommand{\one}{1\!\!1}

\newcommand{\fussend}{\diamond}

\newcommand{\eps}{\varepsilon}

\newcommand{\dali}{\Box}
\newcommand{\choice}[2]{\ifthenelse{\thechoice=1}{#1}{\ifthenelse{\thechoice=2}{#2}{\left\{  \begin{array}{c}

 #1\\
#2\end{array}\right\}  }}}
 
\newcommand{\lchoice}[2]{\ifthenelse{\thechoice=1}{#1}{\ifthenelse{\thechoice=2}{#2}{\left\{  \begin{array}{c}

 #1\\
#2\end{array}\right.}}}
 
\newcommand{\lcsign}{\ifthenelse{\thechoice=1}{+}{\ifthenelse{\thechoice=2}{-}{\pm}}}
 
\newcommand{\lcmsign}{\ifthenelse{\thechoice=1}{-}{\ifthenelse{\thechoice=2}{+}{\mp}}}

\newcommand{\lcconst}{c}
{}
 
\newcommand{\weyl}{\alpha}

\newcommand{\greq}{=_{g}}

\newcommand{\grequiv}{\equiv_{g}}

\newcommand{\grdef}{:=_{g}}

\newcommand{\Greq}{=_{G}}

\newcommand{\Grequiv}{\equiv_{G}}

\newcommand{\Grdef}{:=_{G}}

\newcommand{\Ggreq}{=_{Gg}}

\newcommand{\Greqornot}{=_{(G)}}

\newcommand{\Grequivornot}{\equiv_{(G)}}

\newcommand{\Grdefornot}{:=_{(G)}}

\newcommand{\greqornot}{=_{(g)}}

\newcommand{\grequivornot}{\equiv_{(g)}}

\newcommand{\grdefornot}{:=_{(g)}}

\newcommand{\fatkomma}{\textrm{{\bf ,}}}

\newcommand{\basis}{\boldsymbol{\mf{t}}}

\newcommand{\ip}{\imath}

\newcommand{\Beta}{\textrm{\Large$\beta$}}

\newcommand{\sBeta}{\textrm{\large$\beta$}}

\newcommand{\be}{\bs{b}}

\newcommand{\ce}{\bs{c}}

\newcommand{\Q}{\bs{Q}}

\newcommand{\mm}{\bs{m}\ldots\bs{m}}

\newcommand{\nn}{\bs{n}\ldots\bs{n}}

\newcommand{\kk}{k\ldots k}

\newcommand{\OO}{\bs{\Omega}}

\newcommand{\oo}{\bs{o}}

\newcommand{\tet}{\bs{\theta}}

\newcommand{\Es}{\bs{S}}

\newcommand{\Ce}{\bs{C}}

\newcommand{\lam}{\bs{\lambda}}

\newcommand{\ro}{\bs{\rho}}

\newcommand{\cov}{\textrm{D}_{\tet}}

\newcommand{\feps}{\bs{\eps}}

\newcommand{\qu}{\textrm{Q}_{\tet}}

\newcommand{\dimw}{d_{\textrm{w}}}

\newcommand{\mteta}{\mu(\tet)}

\newcommand{\msig}{d^{\lqn{\hoch{\dimw}}}\sigma}

\newcommand{\msigp}{d^{\hoch{{\scriptscriptstyle \dimw}\lqn{{\scriptscriptstyle -1}}}}\sigma}

\newcommand{\backtilde}{\!\!\tilde{}\,\,}

\title{Da investiga\c{c}\~ao sobre a natureza da luz \`a relatividade especial}

\author{\begin{picture}(0,0)\unitlength=1mm\put(80,40){Notas de aula}\put(80,35){F\'isica contempor\^anea I}
\end{picture} \textbf{Tiago Carvalho Martins}%
\thanks{tiagocm@ufpa.br%
}\emph{\vspace{.5cm}} \emph{}\\
\emph{Faculdade de F\'isica, Universidade Federal do Sul e Sudeste do Par\'a, UNIFESSPA}\\
\emph{68505-080, PA, Brazil\vspace{.5cm}} \emph{}\\}

\maketitle
\begin{abstract}
It is shown as experiments and theories about the nature of light led to the special theory of relativity.
The most important facts for the emergence of the theory proposed by Einstein in 1905 are presented. \noindent \vspace{2cm} \\
\centering{\textbf{Resumo}} \\
\'E mostrado como experimentos e teorias sobre a natureza da luz levaram \`a teoria da relatividade especial.
Os fatos mais importantes para o surgimento da teoria proposta por Einstein em 1905 s\~ao apresentados. 
\noindent \vspace{2cm}
\end{abstract}
}

\Teil{A}{
\newpage 

\tableofcontents{}\newpage\eref

\section{Relatividade especial}

Em 1905, Albert Einstein lan\c{c}ou os fundamentos da teoria da relatividade especial \cite{Artigo.Einstein:1905},
dando origem \`a mec\^anica relativ\'istica,
que em rela\c{c}\~ao \`a mec\^anica cl\'assica, constitui-se numa teoria mais precisa 
e condizente com os resultados experimentais. Aqui ser\~ao discutidos alguns rudimentos dessa teoria. Textos
mais aprofundados podem ser encontrados em \cite{landau,french}.

\section{Natureza, velocidade e momento linear da luz}

Na antiguidade, Pit\'agoras de Samos no s\'eculo VI a.C. acreditava que a luz era formada por part\'iculas e Arist\'oteles de
Estagira no s\'eculo IV a.C. defendia que a luz propagava-se como onda em um meio denominado de \textit{\'eter}.
Na renascen\c{c}a, Robert Hooke em seu livro \textit{Micrographia} (1665)
fala da luz como uma vibra\c{c}\~ao comunicada atrav\'es de um meio \cite{livro.hooke}.
Em 1671, Isaac Newton ap\'os trabalhar com experimentos em \'optica, prop\~oe uma teoria corpuscular para a luz \cite{newton}.
Em 1690, Christiaan Huygens desenvolveu a teoria ondulat\'oria da luz, a partir do estudo da reflex\~ao e refra\c{c}\~ao
\cite{livro.huygens}. Na idade moderna, Thomas Young publica seus trabalhos sobre o fen\^omeno da interfer\^encia da luz
\cite{young}, em 1804. Posteriormente, surgiriam os trabalhos de Augustin-Jean Fresnel sobre a difra\c{c}\~ao, interfer\^encia
e polariza\c{c}\~ao da luz \cite{fresnel}.

Os primeiros c\'alculos para a velocidade da luz foram obtidos 
a partir de observa\c{c}\~oes astron\^omicas. A partir de dados coletados por Ole Christensen R$\varnothing$mer em 1675, referentes ao eclipse de Io (uma das luas de J\'upiter), 
Christiaan Huygens estimou a velocidade da luz \cite{livro.huygens} (algo equivalente a $c=213~333~333~m/s$).
Em 1829, James Bradley estudando o fen\^omeno da \textit{aberra\c{c}\~ao estelar}, encontrou para a luz
uma velocidade $c=304~000~000~m/s$ \cite{bradley}.

\begin{figure}[!h]
\begin{center}
\includegraphics[scale=0.5, bb = 250 50 400 400]{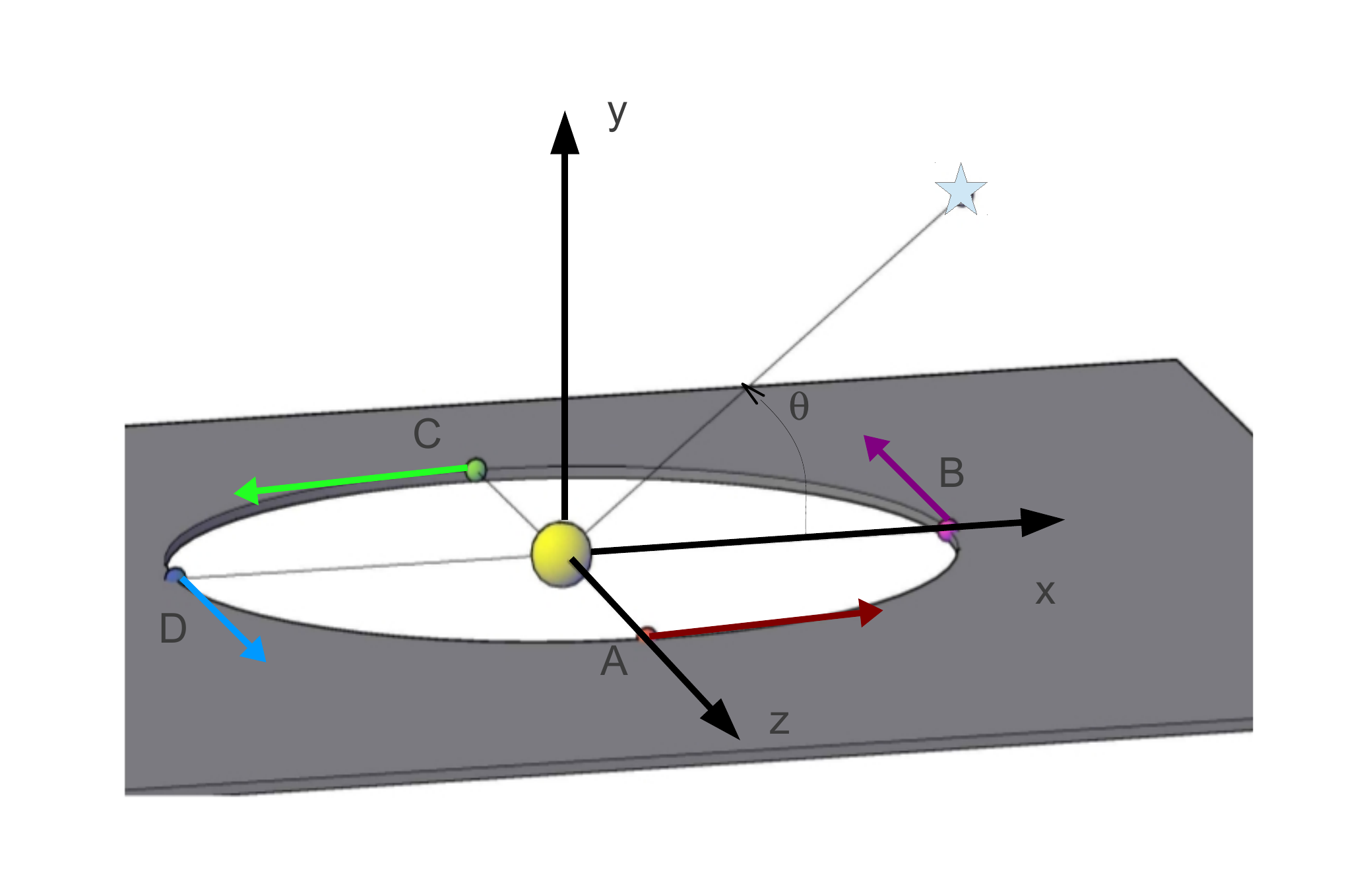}
\end{center}
\caption{James Bradley buscava um m\'etodo de mensurar as dist\^ancias estelares em termos do di\^ametro da \'orbita da Terra.
Ele acreditava que a altitude (expressa pelo \^angulo $\theta$) da estrela $\gamma$-Draconis pareceria maior em B do que em D,
e que as altitudes seriam iguais em A e C. Contudo, ap\'os realizar suas observa\c{c}\~oes, constatou que a altitude era 
m\'axima em A e m\'inima em C, e as altitudes em B e D eram iguais, mas a estrela apresentava desvios para a direita e para
a esquerda, respectivamente. Ele conseguiu explicar esse movimento aparente da estrela em termos do movimento da Terra
em sua \'orbita em torno do Sol.}
\label{fig.aberracao1}
\end{figure}

\begin{figure}[!h]
\begin{center}
\includegraphics[scale=0.5, bb = 250 50 400 500]{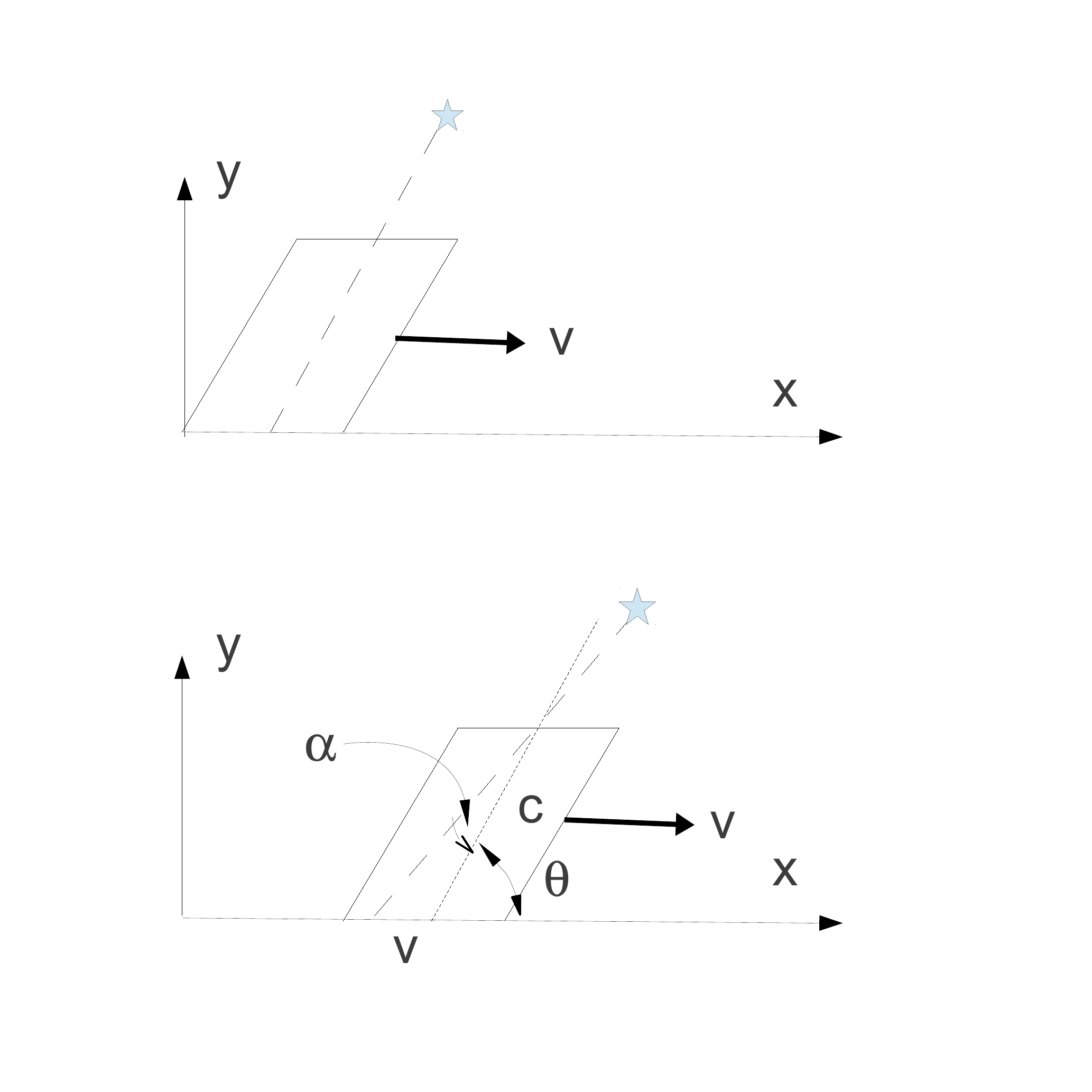}
\end{center}
\caption{Se o telesc\'opio avan\c{c}a em dire\c{c}\~ao \`a estrela (como ocorre no ponto A), a altitude da estrela parece menor.
Utilizando a lei dos senos, vemos que o \^angulo de desvio $\alpha$ (\^angulo de aberra\c{c}\~ao) \'e dado por 
$\alpha\approx\frac{v}{c}sen\theta$. Utilizando o mesmo racioc\'inio para o ponto C, temos $\alpha\approx-\frac{v}{c}sen\theta$,
ou seja, a estrela aparenta ter maior altitude.}
\label{fig.aberracao_FT}
\end{figure}

\begin{figure}[!h]
\begin{center}
\includegraphics[scale=0.5, bb = 250 50 400 500]{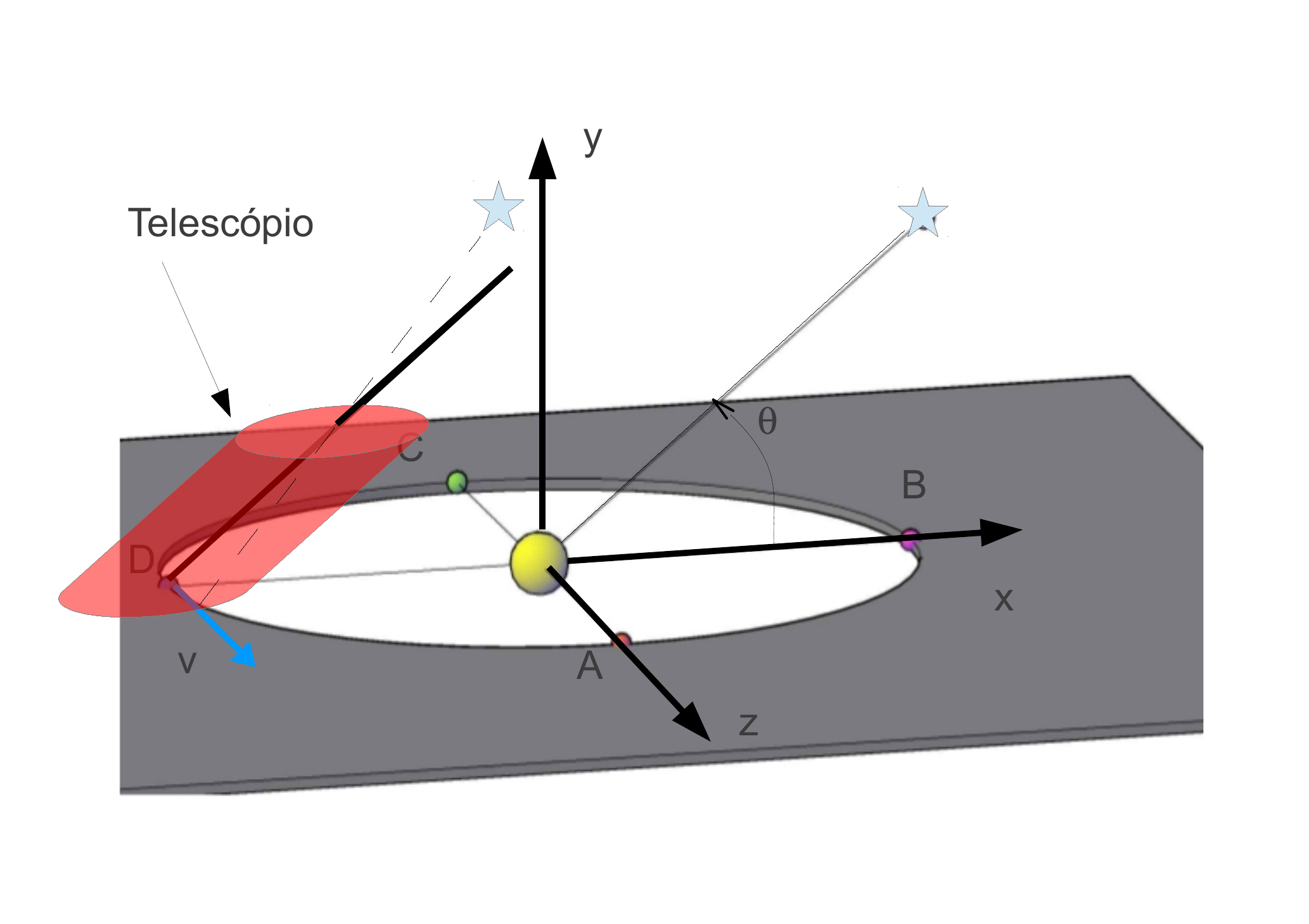}
\end{center}
\caption{Nos pontos B e D a Terra n\~ao se afasta nem se aproxima da estrela, portanto, ela aparece em uma altitude
dada por $\theta$, mas aparenta estar deslocada para a direita no ponto B e para a esquerda no ponto D. Observando a figura,
notamos que esses desvios s\~ao de mesma magnitude, dados por $\alpha\approx\frac{v}{c}$. Esse valor de $\alpha$
obtido foi de cerca de $20,6"$, para uma velocidade orbital da Terra de $v=30~Km/s$, d\'a um valor aproximado 
de $c=300~000~000~m/s$.}
\label{fig.aberracao2}
\end{figure}

Em 1849, Hippolyte Fizeau mediu a velocidade da luz atrav\'es do experimento ilustrado na figura \ref{fig.fizeau1} \cite{fizeau},
obtendo uma velocidade da luz de $c=313~274~304~m/s$.
Em 1962, Foucault refinou esse experimento \cite{foucault}, utilizando espelhos rotativos em vez de uma roda dentada, 
e obteve um melhor resultado: $c=298~000~000~m/s$.

\begin{figure}[!h]
\begin{center}
\includegraphics[scale=0.5, bb = 250 50 400 400]{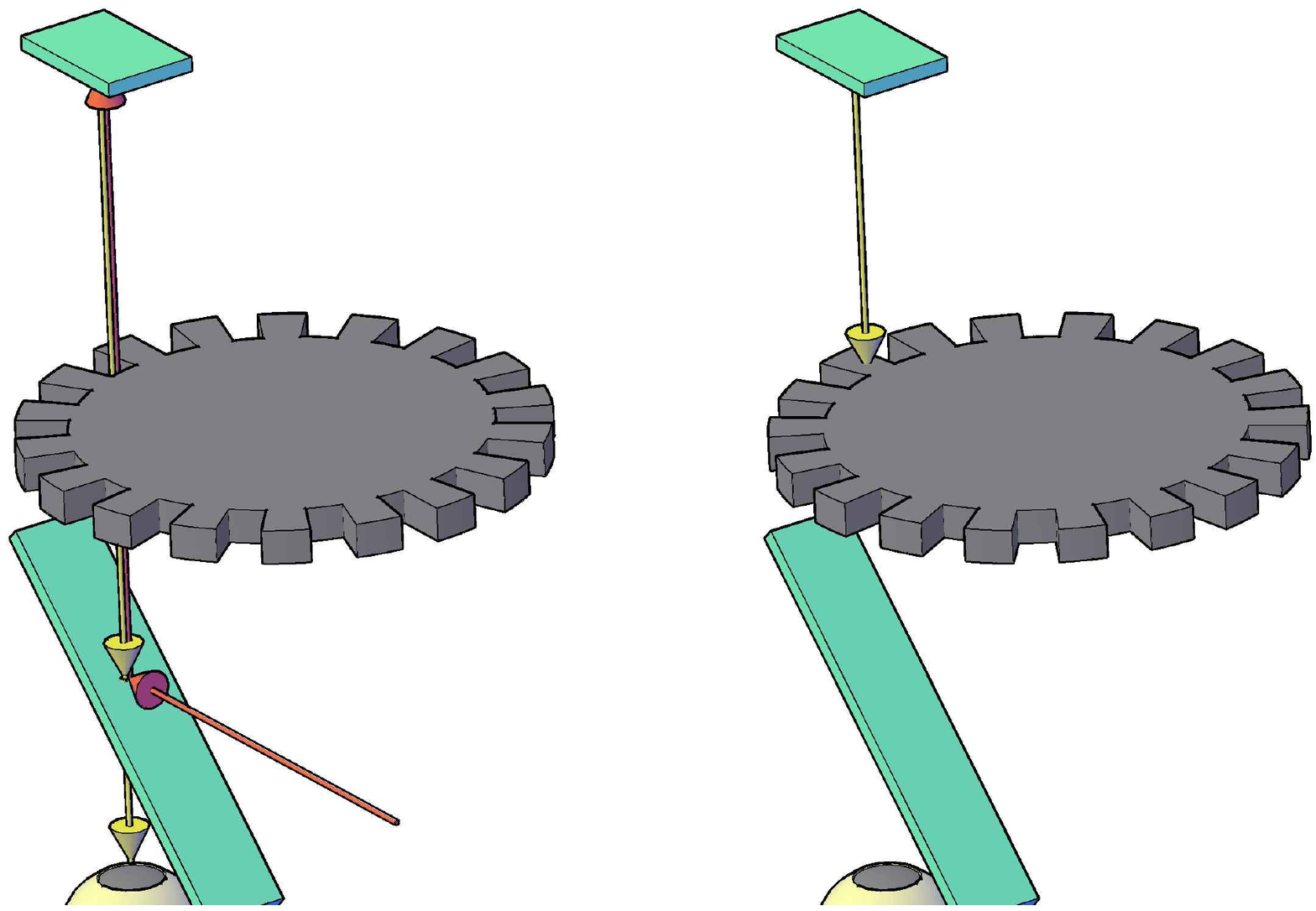}
\end{center}
\caption{Experimento utilizado por Fizeau para medir a velocidade de propaga\c{c}\~ao da luz. Na figura do lado direito
os raios incidentes n\~ao foram representados. Nesse experimento, um feixe de luz \'e parcialmente refletido por um espelho semi-prateado, 
o feixe luminoso refletido atravessa uma das fendas da roda dentada que gira com velocidade angular $\omega=2\pi f$, 
atinge um espelho localizado a uma dist\^ancia $d$ e volta, podendo atravessar, caso
encontre uma fenda; ou n\~ao, caso encontre um dente.}
\label{fig.fizeau1}
\end{figure}

\begin{figure}[!h]
\begin{center}
\includegraphics[scale=0.5]{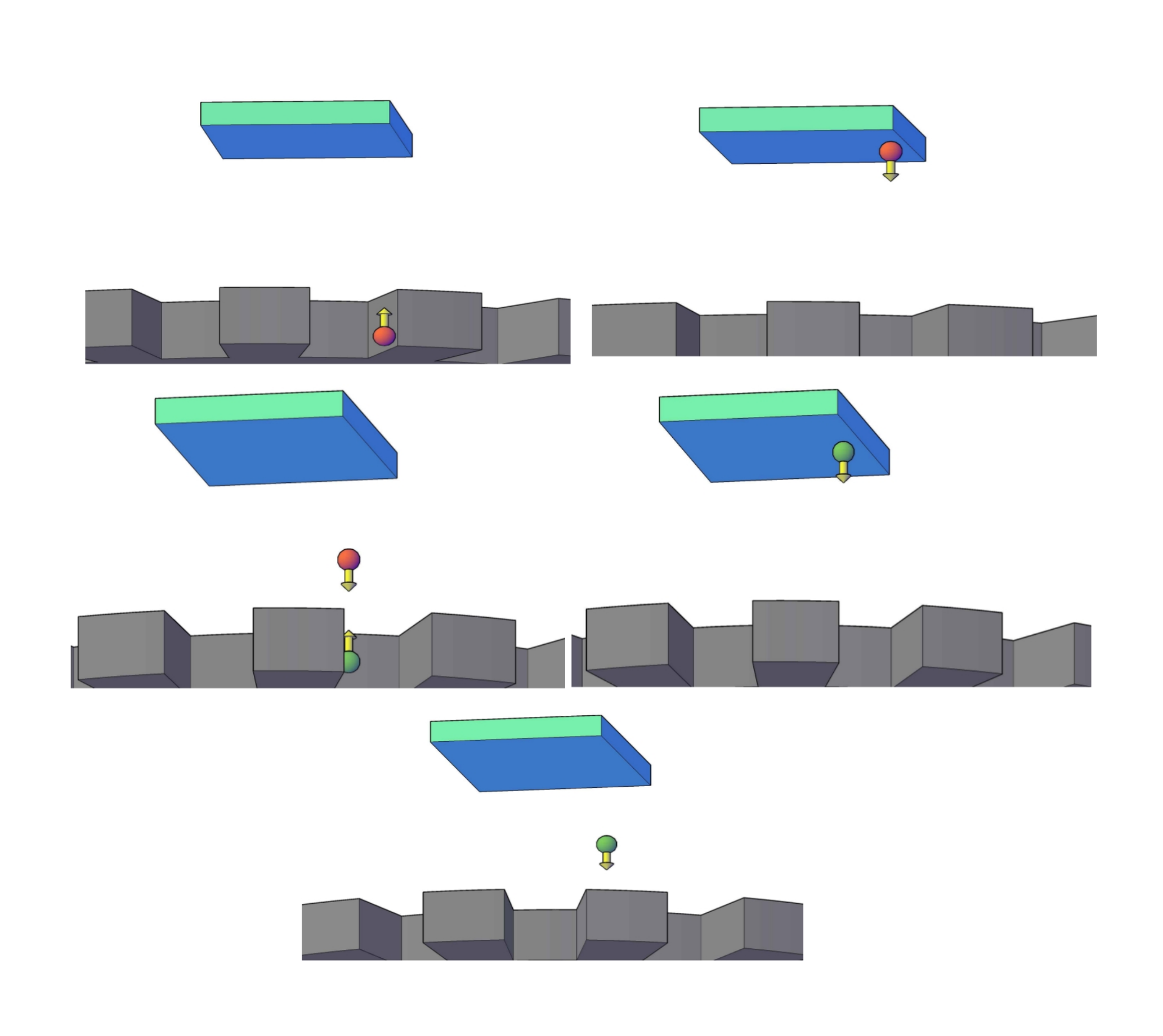}
\end{center}
\caption{Em geral, alguns raios sempre chegam ao olho do observador. Os poucos
casos em que isso n\~ao acontece, ocorrem quando o feixe incidente atravessa uma fenda no exato momento 
em que uma fenda surge e volta no exato momento em que um dente aparece. Nesse caso, o \'ultimo raio que atravessa a fenda 
volta quando a pr\'oxima fenda ainda est\'a na imin\^encia de surgir. 
Dessa forma, se o tempo de ida e volta da luz ($=2d/c$) coindide com
o tempo ($=\theta/\omega$) para a roda girar de um \^angulo $\theta$ 
(igual ao \^angulo central correspondente a uma fenda), ent\~ao,
nenhuma luz \'e vista pelo observador. No experimento de Fizeau, 
o espelho estava fixo a uma dist\^ancia $d=8633~m$, a roda dentada possuia $720~dentes$ e
girava com uma frequ\^encia de $f=12.6~Hz$, quando eclipsou a luz. Logo, a velocidade da luz obtida foi de $c=313~274~304~m/s$.}
\label{fig.figurafizeau}
\end{figure}

Em 1865, James Clerk Maxwell calculou a velocidade da luz teoricamente \cite{maxwell} 
(um c\'alculo usando a nota\c{c}\~ao vetorial \'e apresentado no ap\^endice \ref{calculoc}), utilizando os par\^ametros 
eletrodin\^amicos obtidos por Weber e Kohlrausch \cite{weber}, obtendo $c=310~740~000~m/s$, estando de acordo 
com os resultados experimentais obtidos por Fizeau e Foucault. A posi\c{c}\~ao de destaque assumida 
pela teoria ondulat\'oria da luz em detrimento da teoria corpuscular duraria at\'e 1905, quando Einstein 
prop\^os a ideia de que a luz \'e formada por part\'iculas (posteriormente denominadas de f\'otons). A partir dessa teoria,
ele conseguiu explicar o efeito fotoel\'etico \cite{einsteinefeitofotoeletrico}, estabelecendo que um aumento na 
energia produz um aumento na frequ\^encia do f\'oton, mas n\~ao altera a sua velocidade; sendo $c$, uma
constante fundamental da natureza.

A velocidade de uma onda mec\^anica  \'e sempre definida com rela\c{c}\~ao a um referencial, por exemplo, quando
dizemos que a velocidade
do som \'e de $340~m/s$, estamos nos referindo a velocidade com rela\c{c}\~ao \`as mol\'eculas de ar.
Ent\~ao surge a pergunta: "A velocidade da luz de $c=299~792~458~m/s$ \'e medida em rela\c{c}\~ao a que referencial?" 
A resposta dos f\'isicos a \'epoca de Maxwell \'e a de que existiria um meio, 
denominado de \textit{\'eter lumin\'ifero}, em rela\c{c}\~ao ao qual a luz mover-se-ia com velocidade $c$. Essa teoria
foi derrubada pelo experimento de Michelson e Morley. Em 1905, a teoria da relatividade especial daria uma resposta a esse
problema: $c$ \'e o mesmo para todos os referenciais inerciais, ou seja, n\~ao existe um referencial privilegiado.

Maxwell previu teoricamente que as ondas eletromagn\'eticas al\'em de enegia, transportam momento linear \cite{livro.maxwell}
(um c\'alculo simplificado \'e apresentado no ap\^endice \ref{calculop}),
e portanto, exercem uma press\~ao sobre os objetos em que incidem (press\~ao de radia\c{c}\~ao). Esse fato foi comprovado
posteriormente atrav\'es de experimentos \cite{nichols}. 

Se uma onda eletromagn\'etica incide sobre uma placa condutora, o momento linear $p_e$ exercido sobre um el\'etron
da placa \'e:

\begin{equation} 
p_e = \frac{U_e}{c}.
\end{equation} 
Em que $U_e$ \'e a energia absorvida pelo el\'etron. 
Em termos do f\'oton, podemos dizer que a energia $E_f=hf$ de um f\'oton de momento linear $p_f$ \'e:

\begin{equation} 
E_f = cp_f
\end{equation} 

\begin{figure}[!h]
\begin{center}
\includegraphics[scale=0.4]{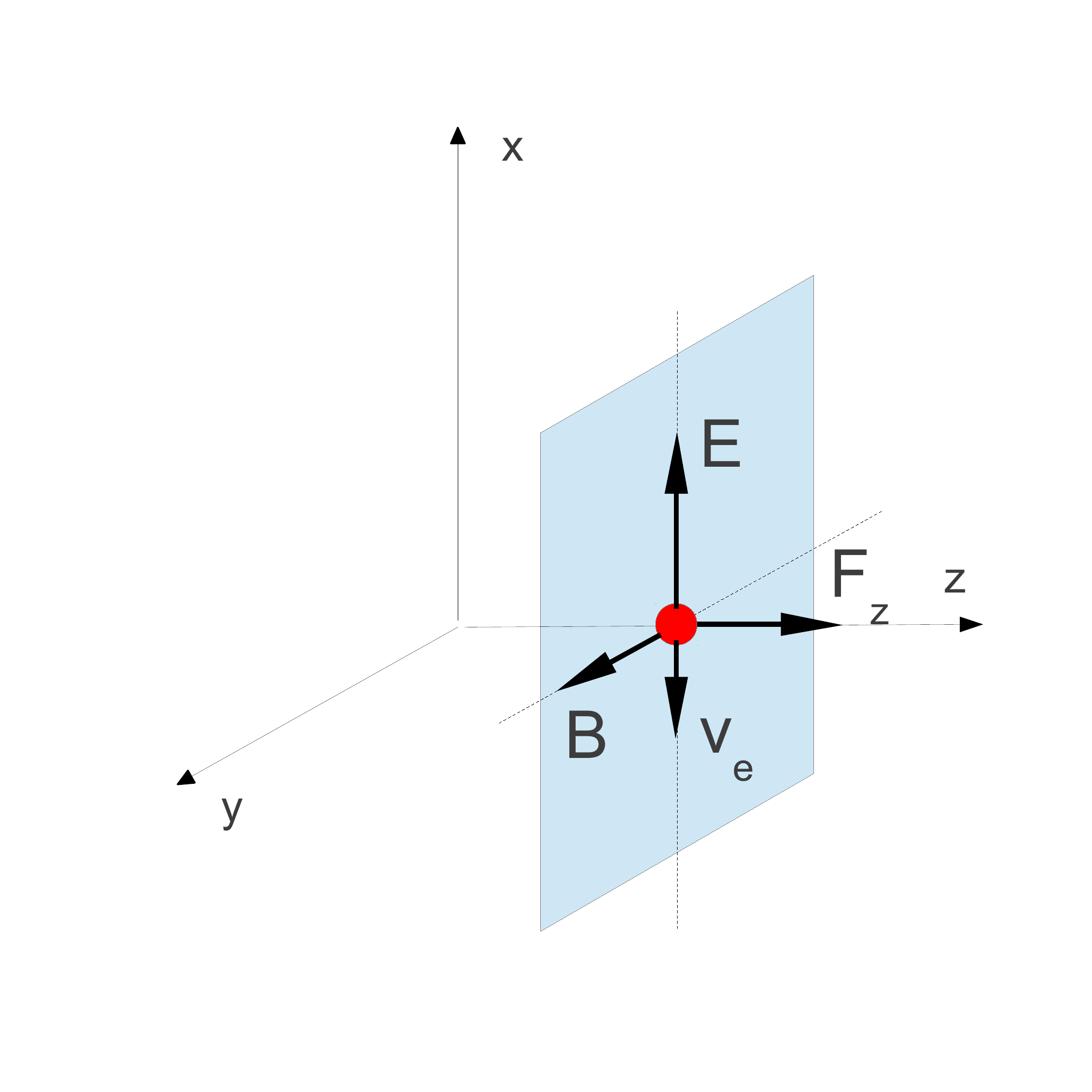}
\end{center}
\caption{Press\~ao de radia\c{c}\~ao: a onda eletromagn\'etica exerce uma for\c{c}a $F_z$ sobre o el\'etron.}
\label{fig.pressaoradiacao}
\end{figure}

\section{Experimento de Michelson-Morley}

Em 1887, o experimento de Michelson-Morley falhou em seu objetivo de tentar comprovar a exist\^encia do 
hipot\'etico \textit{\'eter lumin\'ifero} \cite{michelson}
(o qual seria o referencial absoluto em rela\c{c}\~ao ao qual a velocidade da luz \'e medida),
contudo, a verdadeira import\^ancia desse experimento \'e que ele demonstra 
que a velocidade da luz \'e sempre a mesma para diferentes referenciais inerciais.

\begin{figure}[h!]
\centerline{\includegraphics[scale=0.1]{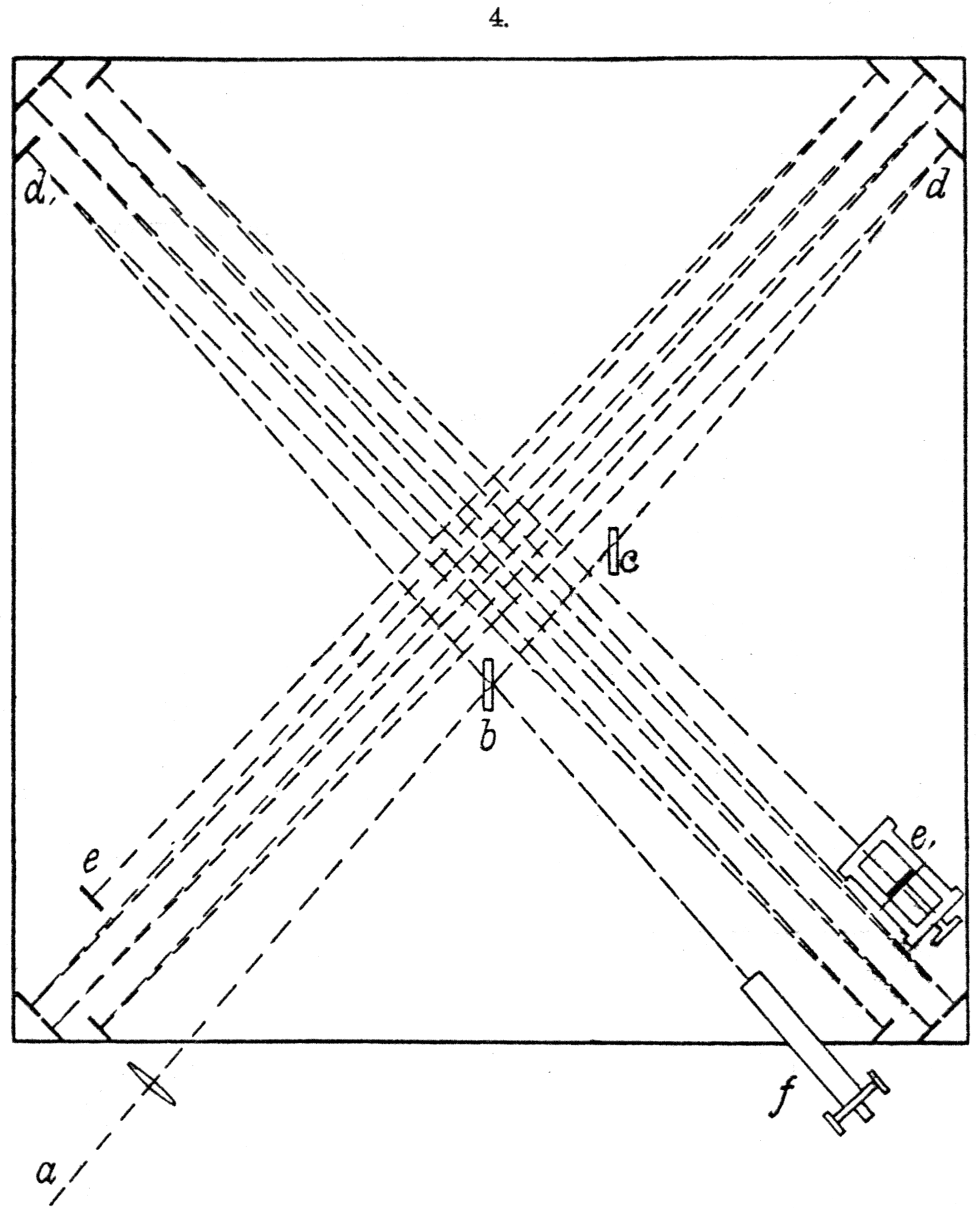}}
\caption{O feixe de luz emitido pela fonte $a$
incide no divisor de luz $b$ (um espelho semi-prateado) e \'e dividido em dois feixes de luz, os quais v\~ao em dire\c{c}\~ao
aos espelhos $d$ e $d'$ ($c$ \'e uma l\^amina de vidro que compensa o fato dos raios oriundos de $b$ n\~ao atravessarem a
mesma espessura de vidro), os feixes de luz sofrem v\'arias reflex\~oes at\'e atingirem os espelhos $e$ e $e'$ e retornam
para $b$ onde s\~ao unidos novamente, indo em dire\c{c}\~ao ao telesc\'opio $f$ 
onde s\~ao observadas as franjas de interfer\^encia
produzidas. No total, cada feixe de luz percorre uma dist\^ancia $l=2\times 10^6\lambda$, entre a sa\'ida e chegada em $b$.
}
\label{fig:interferometro}
\end{figure}

O feixe de luz do interfer\^ometro alinhado com a dire\c{c}\~ao do movimento 
de transla\c{c}\~ao da Terra, 
pela lei de adi\c{c}\~ao de velocidades,
percorreria a dist\^ancia $l$ duas vezes, ora com velocidade $c+v$, ora com velocidade $c-v$ (ver figura \ref{fig.velocidadeparalela}), 
portanto, o tempo $t_1$ gasto nesse percurso \'e:

\begin{equation}
t_1=\frac{l}{c-v}+\frac{l}{c+v}=l\frac{2c}{c^2-v^2}=\frac{2l}{c}\gamma^2
\end{equation}onde $\gamma=\frac{1}{\sqrt{1-\beta^2}}$, com $\beta=v/c=1\times 10^{-4}$ (pois $v\approx 30~Km/s$).

\begin{figure}[!h]
\begin{center}
\includegraphics[scale=0.3, bb = 250 50 400 450]{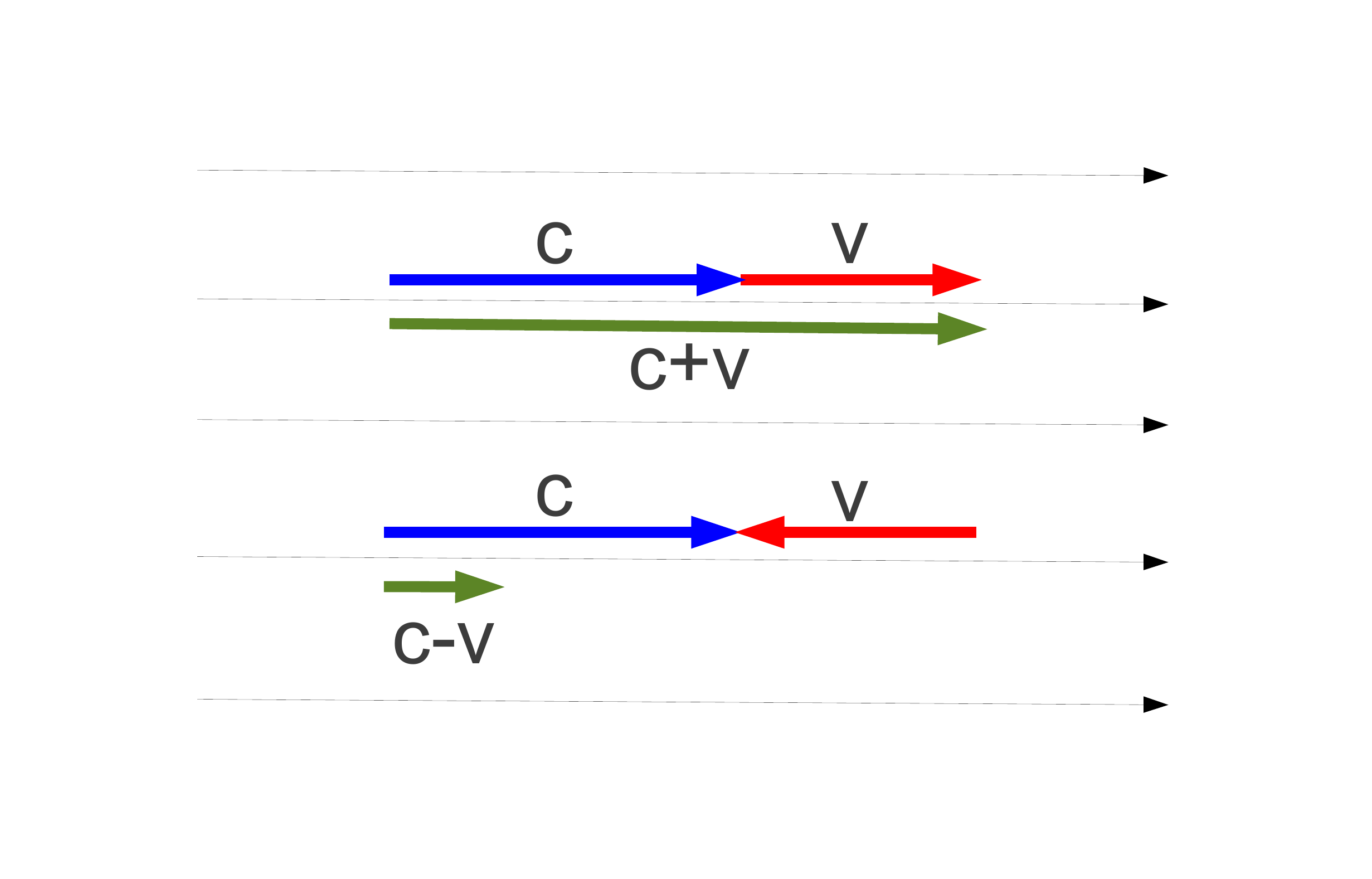}
\end{center}
\caption{As linhas representam o movimento do \'eter. O feixe de luz do interfer\^ometro alinhado com a dire\c{c}\~ao do movimento 
de transla\c{c}\~ao da Terra, 
pela lei de adi\c{c}\~ao de velocidades,
percorreria a dist\^ancia $l$ duas vezes, ora com velocidade $c+v$, ora com velocidade $c-v$.}
\label{fig.velocidadeparalela}
\end{figure}

Por sua vez, se o feixe de luz estiver alinhado perpendicularmente com a dire\c{c}\~ao do movimento de 
transla\c{c}\~ao da Terra, pela lei de adi\c{c}\~ao de velocidades, tem-se 
$c^2=\left(\frac{l}{t_2/2}\right)^2+v^2$ (ver figura \ref{fig.velocidadetransversal}), portanto,
o tempo $t_2$ gasto nesse percurso \'e:

\begin{equation}
t_2=\frac{2l}{c}\gamma
\end{equation}

\begin{figure}[!h]
\begin{center}
\includegraphics[scale=0.3, bb = 250 50 400 400]{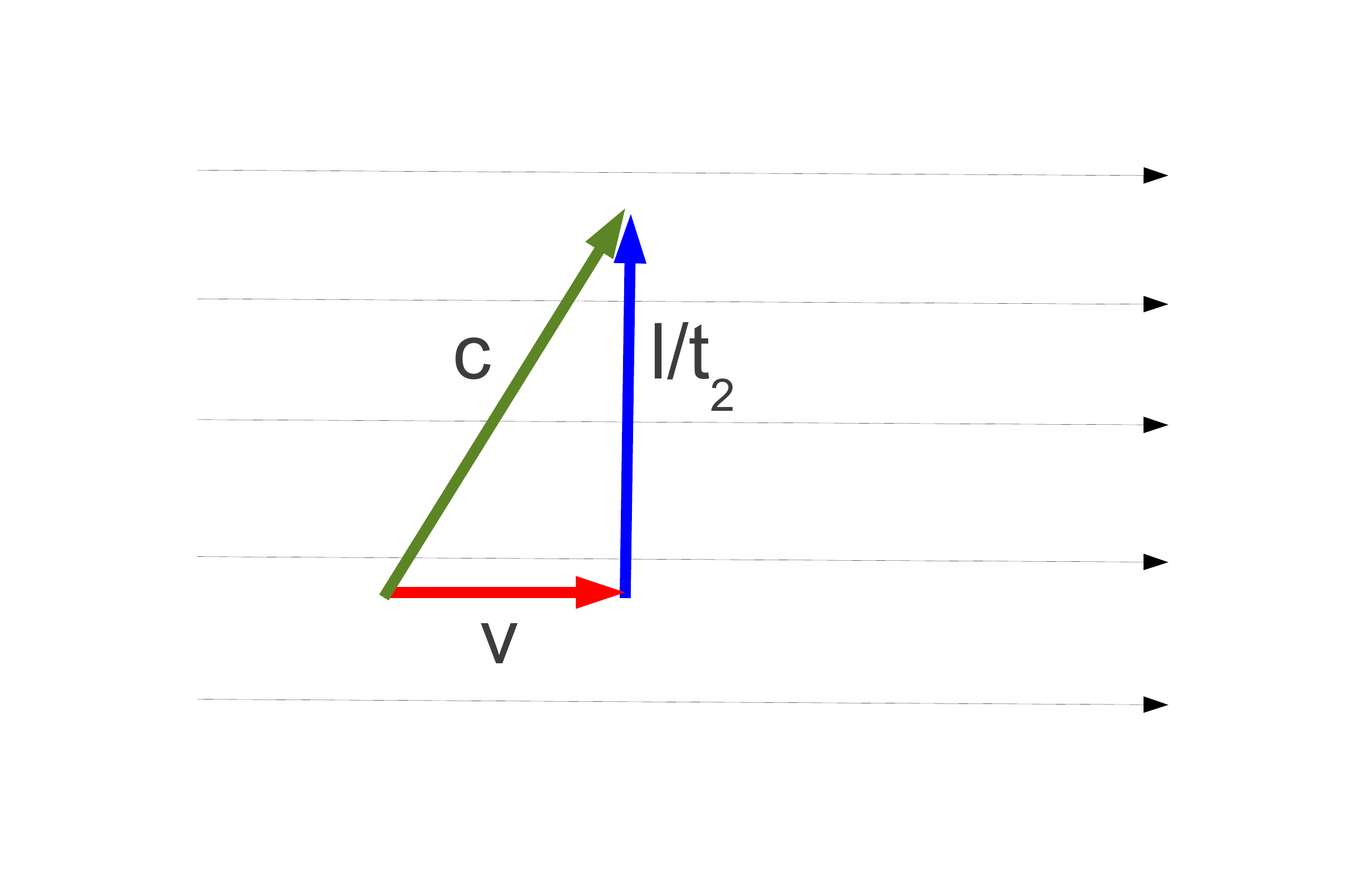}
\end{center}
\caption{se o feixe de luz estiver alinhado perpendicularmente com a dire\c{c}\~ao do movimento de 
transla\c{c}\~ao da Terra, pela lei de adi\c{c}\~ao de velocidades, tem-se 
$c^2=\left(\frac{l}{t_2/2}\right)^2+v^2$.}
\label{fig.velocidadetransversal}
\end{figure}

Dessa forma, o feixe de luz que est\'a alinhado perpendicularmente \`a dire\c{c}\~ao do movimento de transla\c{c}\~ao da Terra 
chega com um atraso de tempo $\Delta t = t_1 - t_2$ com rela\c{c}\~ao ao que est\'a alinhado paralelamente \`a dire\c{c}\~ao de movimento,
dado por:

\begin{equation}
\Delta t=t_1-t_2=\frac{2l}{c}(\gamma^2-\gamma)=\frac{2l}{c}[(1-\beta^2)^{-1}-(1-\beta^2)^{-1/2}]
\label{diferencatempo}
\end{equation}

Como $\beta<<1$, valem as aproxima\c{c}\~oes $(1-\beta^2)^{-1}\approx 1+\beta^2$ e $(1-\beta^2)^{-1/2}\approx 1+\beta^2/2$, 
as quais podem ser obtidas a partir da aproxima\c{c}\~ao linear de $(1+x)^n\approx 1 + n(1+x)$, portanto:

\begin{equation}
\Delta t\approx\frac{l}{c}\beta^2
\end{equation}

Como os feixes possuem uma mesma frequ\^encia, ocorre interfer\^encia. 
O deslocamento $\delta$ produzido nas franjas de interfer\^encia \'e dado pela raz\~ao entre a diferen\c{c}a de percurso 
$c\Delta t$ e o comprimento de onda $\lambda$ \'e:

\begin{equation}
\delta_1=\frac{l}{\lambda}\beta^2
\end{equation}

Se o interfer\^ometro for girado de $\pi/2~rad$, tem-se:

\begin{equation}
\delta_2=-\frac{l}{\lambda}\beta^2
\end{equation}

Como $v\approx 30~Km/s$, e $c\approx 300~000~Km/s$, ent\~ao $\beta=10^{-4}$. Dessa forma,
o deslocamento de franjas total m\'aximo obtido deve ser:

\begin{equation}
\delta=2\frac{l}{\lambda}\beta^2=0,04~franjas
\end{equation}

Contudo, nenhum deslocamento de franjas significativo foi detectado pelo experimento.

\section{Hip\'otese de contra\c{c}\~ao de Lorentz-FitzGerald}

Em trabalhos de 1889 e 1892, respectivamente, George FitzGerald e Hendrik Lorentz
propuseram a hip\'otese de contra\c{c}\~ao de Lorentz-FitzGerald \cite{fitzgerald,lorentz1}, a fim de conciliar a 
exist\^encia do \'eter com uma explica\c{c}\~ao para o resultado nulo do experimento de Michelson-Morley. Segundo essa
hip\'otese, a dimens\~ao de um objeto na dire\c{c}\~ao do movimento em rela\c{c}\~ao ao \'eter seria contraida. Isso
aconteceria com os bra\c{c}os do interfer\^ometro, portanto, de \ref{diferencatempo} ter-se-ia:

\begin{equation}
\Delta t=t_1-t_2=\frac{2l_1}{c}\gamma^2-\frac{2l_2}{c}\gamma \Rightarrow l_1=\frac{l_2}{\gamma}.
\end{equation}

Um experimento similar ao interfer\^ometro de Michelson-Morley foi proposto por Kennedy e Thorndike \cite{KennedyThorndike},
mas com bra\c{c}os de comprimentos diferentes. Esse experimento mostrou-se incompat\'ivel com a hip\'otese de contra\c{c}\~ao 
de Lorentz-FitzGerald, pois necessitava levar em conta o efeito de dilata\c{c}\~ao temporal, o qual \'e previsto pela
teoria da relatividade especial.

\section{Transforma\c{c}\~ao galileana versus transforma\c{c}\~ao de Lorentz}

\subsection{Transforma\c{c}\~ao galileana na mec\^anica}

Todo fen\^omeno f\'isico \'e descrito em rela\c{c}\~ao a um referencial. O referencial \'e o sistema em rela\c{c}\~ao 
ao qual s\~ao feitas as medidas das coordenadas espaciais 
$x$, $y$ e $z$ que indicam a posi\c{c}\~ao de uma part\'icula, assim como, do tempo $t$ decorrido. 

Um referencial \'e dito inercial quando o movimento livre das part\'iculas (movimento com resultante das for\c{c}as externas
nula) ocorre com velocidade constante. Dado um referencial inercial, qualquer outro referencial movendo-se com velocidade
constante em rela\c{c}\~ao a ele tamb\'em ser\'a inercial.

Um conjunto de equa\c{c}\~oes que descreve um fen\^omeno f\'isico em fun\c{c}\~ao das coordenadas $(t,x,y,z)$ 
no re- ferencial inercial 
$S$ pode ser reescrito em fun\c{c}\~ao das coordenadas $(t',x',y',z')$ no referencial inercial $S'$. 
Se essa transforma\c{c}\~ao de coordenadas preserva
a forma das equa\c{c}\~oes (\textit{leis f\'isicas}), 
dizemos que essas leis f\'isicas s\~ao \textit{invariantes} sob tal transforma\c{c}\~ao. 

Um fen\^omeno f\'isico ao ser descrito por um sistema de coordenadas quadridimensional $(t,x,y,z)$
\'e chamado de \textit{evento}. O espa\c{c}o vetorial 4-dimensional formado pelos pontos $(t,x,y,z)$ \'e chamado de 
\textit{espa\c{c}o de Minkowski}.

Dados um referencial inercial $S'$ que se move com velocidade constante $\vec{v}$ em rela\c{c}\~ao ao referencial inercial $S$, 
a mudan\c{c}a de coordenadas
de $S$ para $S'$ \'e chamada de \textit{transforma\c{c}\~ao galileana}, dada por:
\begin{equation}
\vec{r}'=\vec{r}-\vec{v}t,~~~~~~~~~~~~~~~~~~~~t'=t \label{transformacaogalileana}
\end{equation}
As velocidades $\vec{u}(t)$ em $S$ e $\vec{u'}(t)$ em $S'$, 
para uma part\'icula de posi\c{c}\~oes $\vec{r}(t)$ em $S$ e $\vec{r'}(t)$ em $S'$, s\~ao dadas por:
\begin{equation}
\vec{u}=\frac{d\vec{r}}{dt},~~~~~~~~~~\vec{u'}=\frac{d\vec{r'}}{dt'}=\frac{d\vec{r'}}{dt}\frac{dt}{dt'}=\frac{d\vec{r'}}{dt}
=\frac{d(\vec{r}-\vec{v}t)}{dt}=\frac{d\vec{r}}{dt}-\vec{v}=\vec{u}-\vec{v} \label{adicaodevelocidade}
\end{equation}
A seu turno, as acelera\c{c}\~oes $\vec{a}(t)$ em $S$ e $\vec{a'}(t)$ em $S'$ s\~ao:
\begin{equation}
\vec{a}=\frac{d\vec{u}}{dt},~~~~~
\vec{a'}=\frac{d\vec{u'}}{dt'}=\frac{d\vec{u'}}{dt}\frac{dt}{dt'}=\frac{d\vec{u'}}{dt}
=\frac{d(\vec{u}-\vec{v})}{dt}=\frac{d\vec{u}}{dt} ~~~~~\Rightarrow \vec{a'}=\vec{a}
\end{equation}
O movimento de uma part\'icula de massa $m$ \'e descrita na mec\^anica atrav\'es da segunda lei de Newton, que nos referenciais $S$
e $S'$ apresentam as formas:
\begin{equation}
\vec{F_R}=m\vec{a},~~~~~~~~~~\vec{F'_R}=m\vec{a'}~~~~~\Rightarrow \vec{F'_R}=\vec{F_R}
\end{equation}
Isso significa que a part\'icula apresenta as mesmas equa\c{c}\~oes de movimento tanto em $S$ quanto em $S'$.
Em outras palavras, \textit{as equa\c{c}\~oes de movimento da mec\^anica s\~ao invariantes sob uma transforma\c{c}\~ao galileana}.
Al\'em disso, como todos os referenciais inerciais apresentam as mesmas
equa\c{c}\~oes de movimento, \'e poss\'ivel afirmar que \textit{n\~ao existe um referencial inercial preferencial para as equa\c{c}\~oes
de movimento da mec\^anica}. Dessa forma, n\~ao h\'a nenhum experimento mec\^anico que se possa realizar dentro de um laborat\'orio
atrav\'es do qual possamos descobrir a velocidade absoluta do mesmo, medida em rela\c{c}\~ao a um referencial de repouso absoluto, 
ou seja, \textit{na mec\^anica n\~ao existem referenciais de repouso absoluto}. Assim, a relatividade galileana estabelece
que o estado de repouso ou movimento \'e relativo, pois, s\'o \'e poss\'ivel afirmar que um corpo possui certa velocidade
ou est\'a em repouso em rela\c{c}\~ao a um certo referencial.

Os resultados do experimento de Michelson-Morley demonstaram que para a luz tamb\'em n\~ao existe um referencial privilegiado,
portanto, espera-se que as equa\c{c}\~oes de Maxwell, que s\~ao as leis que descrevem o comportamento dos 
campos eletromagn\'eticos, tamb\'em sejam invariantes sob uma transforma\c{c}\~ao galileana.

\subsection{Transforma\c{c}\~ao galileana no eletromagnetismo}

Considerando uma fun\c{c}\~ao vetorial da posi\c{c}\~ao e do tempo $\vec{F}(t,x,y,z)$ e 
reescrevendo (\ref{transformacaogalileana}) na forma escalar:
\begin{eqnarray}
x'=x-v_x t \nonumber \\
y'=y-v_y t \nonumber \\
z'=x-v_z t \nonumber \\
t'=t \label{transformacaogalileana1}
\end{eqnarray}
O c\'alculo de $\frac{\partial}{\partial t}$, $\frac{\partial}{\partial x}$, $\frac{\partial}{\partial y}$ e
$\frac{\partial}{\partial z}$ \'e feito pela regra da cadeia:
\begin{eqnarray}
\frac{\partial}{\partial t}
&=&
\frac{\partial t'}{\partial t}\frac{\partial}{\partial t'}
+\frac{\partial x'}{\partial t}\frac{\partial}{\partial x'}
+\frac{\partial y'}{\partial t}\frac{\partial}{\partial y'}
+\frac{\partial z'}{\partial t}\frac{\partial}{\partial z'} \nonumber \\
&=&
\frac{\partial}{\partial t'}
-v_x\frac{\partial}{\partial x'}
-v_y\frac{\partial}{\partial y'}
-v_z\frac{\partial}{\partial z'} \nonumber \\
&=&
\frac{\partial}{\partial t'}
-\vec{v}\cdot\nabla' \label{partialtlinha}
\end{eqnarray}
\begin{equation}
\frac{\partial}{\partial x}
=
\frac{\partial x'}{\partial x}\frac{\partial}{\partial x'}
+\frac{\partial y'}{\partial x}\frac{\partial}{\partial y'}
+\frac{\partial z'}{\partial x}\frac{\partial}{\partial z'}
+\frac{\partial t'}{\partial x}\frac{\partial}{\partial t'}
=
\frac{\partial}{\partial x'}
\end{equation}
\begin{equation}
\frac{\partial}{\partial y}
=
\frac{\partial x'}{\partial x}\frac{\partial}{\partial x'}
+\frac{\partial y'}{\partial x}\frac{\partial}{\partial y'}
+\frac{\partial z'}{\partial x}\frac{\partial}{\partial z'}
+\frac{\partial t'}{\partial x}\frac{\partial}{\partial t'}
=
\frac{\partial}{\partial y'}
\end{equation}
\begin{equation}
\frac{\partial}{\partial z}
=
\frac{\partial x'}{\partial x}\frac{\partial}{\partial x'}
+\frac{\partial y'}{\partial x}\frac{\partial}{\partial y'}
+\frac{\partial z'}{\partial x}\frac{\partial}{\partial z'}
+\frac{\partial t'}{\partial x}\frac{\partial}{\partial t'}
=
\frac{\partial}{\partial z'}
\end{equation}
Portanto:
\begin{equation}
\nabla'
=(\frac{\partial}{\partial x'},\frac{\partial}{\partial y'},\frac{\partial}{\partial z'})
=(\frac{\partial}{\partial x},\frac{\partial}{\partial y},\frac{\partial}{\partial z})
=\nabla \label{nablalinha}
\end{equation}
Como $\vec{v}$ \'e constante, ent\~ao, $\nabla'\cdot\vec{v}=0$ e $(\vec{F}\cdot\nabla')\vec{v}=0$, os quais podem ser
substitu\'idos na identidade vetorial a seguir:
\begin{eqnarray}
\nabla'\times(\vec{v}\times\vec{F})=\vec{v}(\nabla'\cdot\vec{F})-\vec{F}(\nabla'\cdot\vec{v})
+(\vec{F}\cdot\nabla')\vec{v}-(\vec{v}\cdot\nabla')\vec{F} \nonumber \\
\nabla'\times(\vec{v}\times\vec{F})=\vec{v}(\nabla'\cdot\vec{F})-(\vec{v}\cdot\nabla')\vec{F}
\Rightarrow -(\vec{v}\cdot\nabla')\vec{F}=\nabla'\times(\vec{v}\times\vec{F})-\vec{v}(\nabla'\cdot\vec{F}) \label{identidade1}
\end{eqnarray}

Os vetores que aparecem nas equa\c{c}\~oes (\ref{faraday}), (\ref{ampere}), (\ref{gausse}) e (\ref{gaussh}), 
com rela\c{c}\~ao ao referencial inercial $S$, s\~ao fun\c{c}\~oes da posi\c{c}\~ao $\vec{r}$ e do tempo $t$:
$\vec{E}(\vec{r},t)$, $\vec{H}(\vec{r},t)$, $\vec{D}(\vec{r},t)$ e $\vec{B}(\vec{r},t)$.
Contudo, com rela\c{c}\~ao ao referencial inercial $S'$, esses vetores s\~ao fun\c{c}\~oes da posi\c{c}\~ao $\vec{r'}$ e do tempo $t'$:
$\vec{E'}(\vec{r'},t')$, $\vec{H'}(\vec{r'},t')$, $\vec{D'}(\vec{r'},t')$ e $\vec{B'}(\vec{r'},t')$.

Para realizar a transforma\c{c}\~ao galileana da lei de Faraday, substituimos (\ref{nablalinha}) e (\ref{partialtlinha})
em (\ref{faraday}), e em seguida utilizamos (\ref{identidade1}), obtendo:
\begin{eqnarray}
\nabla'\times\vec{E}=-\frac{\partial\vec{B}}{\partial t'}-\nabla'\times(\vec{v}\times\vec{B})+\vec{v}(\nabla'\cdot\vec{B})
\nonumber \\
\Rightarrow 
\nabla'\times (\vec{E}+\vec{v}\times\vec{B})=-\frac{\partial\vec{B}}{\partial t'}+\vec{v}(\nabla'\cdot\vec{B})
\label{faradaylinha}
\end{eqnarray}

Por sua vez, para realizar a transforma\c{c}\~ao galileana da lei de Ampere, substituimos (\ref{nablalinha}) e (\ref{partialtlinha})
em (\ref{ampere}), e em seguida utilizamos (\ref{identidade1}), obtendo:
\begin{eqnarray}
\nabla'\times\vec{H}=
\vec{J}+\frac{\partial\vec{D}}{\partial t'}+\nabla'\times(\vec{v}\times\vec{D})-\vec{v}(\nabla'\cdot\vec{D})
\nonumber \\
\Rightarrow 
\nabla'\times (\vec{H}-\vec{v}\times\vec{D})=
\vec{J}+\frac{\partial\vec{D}}{\partial t'}-\vec{v}(\nabla'\cdot\vec{D})
\label{amperelinha}
\end{eqnarray}

Al\'em disso, para realizar a transforma\c{c}\~ao galileana da equa\c{c}\~ao da continuidade, 
subs- tituimos (\ref{nablalinha}) e (\ref{partialtlinha})
em (\ref{continuidade}), obtendo:
\begin{eqnarray}
\nabla'\times\vec{J}&=&
-\frac{\partial\rho}{\partial t'}+\vec{v}\cdot(\nabla'\rho)
\Rightarrow 
\nabla'\times\vec{J}=
-\frac{\partial\rho}{\partial t'}
+\vec{v}\cdot(\frac{\partial\rho}{\partial x'},\frac{\partial\rho}{\partial y'},\frac{\partial\rho}{\partial z'})
\nonumber \\
&=&
-\frac{\partial\rho}{\partial t'}
+v_x\frac{\partial\rho}{\partial x'}+v_y\frac{\partial\rho}{\partial y'}+v_z\frac{\partial\rho}{\partial z'}
=
-\frac{\partial\rho}{\partial t'}
+\frac{\partial (v_x\rho)}{\partial x'}
+\frac{\partial (v_y\rho)}{\partial y'}+\frac{\partial (v_z\rho)}{\partial z'} \nonumber \\
&=&
-\frac{\partial\rho}{\partial t'}
+\nabla'(\rho\vec{v})
\Rightarrow \nabla'\times(\vec{J}-\rho\vec{v})=-\frac{\partial\rho}{\partial t'} \label{continuidadelinha}
\end{eqnarray}
Analisando (\ref{faradaylinha}), (\ref{amperelinha}) e (\ref{continuidadelinha}), conclu\'imos que para preservar a forma
das equa\c{c}\~oes (\ref{faraday}), (\ref{ampere}), (\ref{gausse}) e (\ref{gaussh}) sob uma transforma\c{c}\~ao galileana de um
referencial inercial $S$ para um referencial inercial $S'$, devemos fazer:
\begin{eqnarray}
\vec{E'}=\vec{E}+\vec{v}\times\vec{B} \label{elinha}  \\
\vec{D'}=\vec{D}  \label{dlinha} \\
\vec{H'}=\vec{H}-\vec{v}\times\vec{D} \label{hlinha}  \\
\vec{B'}=\vec{B} \label{blinha} \\
\vec{J'}=\vec{J}-\rho\vec{v} \label{jlinha} \\
\rho'=\rho \label{rholinha}
\end{eqnarray}

Multiplicando (\ref{elinha}) por $\varepsilon_0$ e fazendo $\vec{B}=\mu_0\vec{H}$ e, 
multiplicando (\ref{hlinha}) por $\mu_0$ e fazendo $\vec{D}=\varepsilon_0\vec{E}$, ser\~ao obtidas duas equa\c{c}\~oes
nas quais (\ref{velocidadedaluz}) \'e substitu\'ida, resultando em:
\begin{eqnarray}
\varepsilon_0\vec{E'}=\varepsilon_0\vec{E}+\varepsilon_0\mu_0\vec{v}\times\vec{H} 
\Rightarrow \vec{D'}=\vec{D}+\frac{1}{c^2}\vec{v}\times\vec{H} \label{dlinha1} \\
\mu_0\vec{H'}=\mu_0\vec{H}-\mu_0\varepsilon_0\vec{v}\times\vec{H} 
\Rightarrow \vec{B'}=\vec{B}-\frac{1}{c^2}\vec{v}\times\vec{E} \label{blinha1}
\end{eqnarray}
A \'unica maneira de (\ref{dlinha1}) e (\ref{blinha1}) tornarem-se iguais, 
respectivamente, a (\ref{dlinha}) e (\ref{blinha}), seria fazendo $c\rightarrow\infty$. Como sabemos que a velocidade
da luz \'e finita, as equa\c{c}\~oes de Maxwell n\~ao s\~ao invariantes sob uma transforma\c{c}\~ao galileana.

Em 1904, Lorentz apresentou a forma correta para as transforma\c{c}\~oes sofridas pelas equa\c{c}\~oes de Maxwell 
\cite{lorentz2}, quando passam do referencial $S$ para o referencial $S'$. Essas transforma\c{c}\~oes foram denominadas de
transforma\c{c}\~oes de Lorentz.

\subsection{Transforma\c{c}\~ao de Lorentz}

Dados dois referenciais $S$ e $S'$ que se movem um em rela\c{c}\~ao ao outro com velocidade $v$ ao longo do eixo $x$,
uma generaliza\c{c}\~ao da transforma\c{c}\~ao galileana, conhecida como transforma\c{c}\~ao de Lorentz, pode ser obtida por:
\begin{eqnarray}
x'=\gamma(x-vt) \nonumber \\
y'=y \nonumber \\
z'=z \nonumber \\
t'=\gamma(t-vx/c^2) 
\label{lorentz}
\end{eqnarray}

O c\'alculo de $\frac{\partial}{\partial t}$, $\frac{\partial}{\partial x}$, $\frac{\partial}{\partial y}$ e
$\frac{\partial}{\partial z}$ \'e feito pela regra da cadeia:
\begin{equation}
\frac{\partial}{\partial x}
=
\frac{\partial x'}{\partial x}\frac{\partial}{\partial x'}
+\frac{\partial y'}{\partial x}\frac{\partial}{\partial y'}
+\frac{\partial z'}{\partial x}\frac{\partial}{\partial z'}
+\frac{\partial t'}{\partial x}\frac{\partial}{\partial t'}
=
\gamma\frac{\partial}{\partial x'}
-\frac{\gamma v}{c^2}\frac{\partial}{\partial t'}
\end{equation}

\begin{equation}
\frac{\partial}{\partial y}
=
\frac{\partial x'}{\partial y}\frac{\partial}{\partial x'}
+\frac{\partial y'}{\partial y}\frac{\partial}{\partial y'}
+\frac{\partial z'}{\partial y}\frac{\partial}{\partial z'}
+\frac{\partial t'}{\partial y}\frac{\partial}{\partial t'}
=
\frac{\partial}{\partial y'}
\end{equation}

\begin{equation}
\frac{\partial}{\partial z}
=
\frac{\partial x'}{\partial z}\frac{\partial}{\partial x'}
+\frac{\partial y'}{\partial z}\frac{\partial}{\partial y'}
+\frac{\partial z'}{\partial z}\frac{\partial}{\partial z'}
+\frac{\partial t'}{\partial z}\frac{\partial}{\partial t'}
=
\frac{\partial}{\partial z'}
\end{equation}

\begin{equation}
\frac{\partial}{\partial t}
=
\frac{\partial x'}{\partial t}\frac{\partial}{\partial x'}
+\frac{\partial y'}{\partial t}\frac{\partial}{\partial y'}
+\frac{\partial z'}{\partial t}\frac{\partial}{\partial z'}
+\frac{\partial t'}{\partial t}\frac{\partial}{\partial t'}
=
\gamma\frac{\partial}{\partial t'} -\gamma v\frac{\partial}{\partial x'}
\end{equation}

Reescrevendo a lei de Gauss do campo magn\'etico (\ref{gaussh}) e utilizando as derivadas parciais obtidas anteriomente, temos: 
\begin{eqnarray}
\nabla\cdot\vec{B}
=\frac{\partial B_x}{\partial x}+\frac{\partial B_y}{\partial y}+\frac{\partial B_z}{\partial z}=0 \nonumber \\
\Rightarrow 
\gamma\frac{\partial B_x}{\partial x'}-\gamma\frac{v}{c^2}\frac{\partial B_x}{\partial t'}
+\frac{\partial B_y}{\partial y'}+\frac{\partial B_z}{\partial z'}=0 \nonumber \\
\Rightarrow 
\gamma^2v\frac{\partial B_x}{\partial x'}=\gamma^2\frac{v^2}{c^2}\frac{\partial B_x}{\partial t'}
-\gamma v\frac{\partial B_y}{\partial y'}-\gamma v\frac{\partial B_z}{\partial z'}
\end{eqnarray}

Al\'em disso, reescrevendo a lei de Faraday (\ref{faraday}) e utilizando as derivadas parciais obtidas anteriomente, temos: 
\begin{eqnarray}
\frac{\partial E_z}{\partial y}-\frac{\partial E_y}{\partial z}=-\frac{\partial B_x}{\partial t} \nonumber \\
\frac{\partial E_z}{\partial y'}-\frac{\partial E_y}{\partial z'}
=-\gamma\frac{\partial B_x}{\partial t'} +\gamma v\frac{\partial B_x}{\partial x'} \nonumber \\
\gamma\frac{\partial E_z}{\partial y'}-\gamma\frac{\partial E_y}{\partial z'}
=-\gamma^2\frac{\partial B_x}{\partial t'} +\gamma^2 v\frac{\partial B_x}{\partial x'} \nonumber \\
\gamma\frac{\partial E_z}{\partial y'}-\gamma\frac{\partial E_y}{\partial z'}
=-\gamma^2\frac{\partial B_x}{\partial t'} +\gamma^2\frac{v^2}{c^2}\frac{\partial B_x}{\partial t'}
-\gamma v\frac{\partial B_y}{\partial y'}-\gamma v\frac{\partial B_z}{\partial z'}\nonumber \\
\frac{\partial }{\partial y'}(\gamma E_z+\gamma vB_y)-\frac{\partial }{\partial z'}(\gamma E_y-\gamma vB_z)
=-\gamma^2(1-\frac{v^2}{c^2})\frac{\partial B_x}{\partial t'} \label{maxwelllorentz1}
\end{eqnarray}

\begin{eqnarray}
\frac{\partial E_x}{\partial z}-\frac{\partial E_z}{\partial x}=-\frac{\partial B_y}{\partial t} \nonumber \\
\frac{\partial E_x}{\partial z'}-\gamma\frac{\partial E_z}{\partial x'}+\frac{\gamma v}{c^2}\frac{\partial E_z}{\partial t'}=
-\gamma\frac{\partial B_y}{\partial t'} +\gamma v\frac{\partial B_y}{\partial x'} \nonumber \\
\frac{\partial E_x}{\partial z'}
-\frac{\partial }{\partial x'}(\gamma E_z+\gamma vB_y)=
-\frac{\partial }{\partial t'} (\gamma B_y+\frac{\gamma v}{c^2}E_z) \label{maxwelllorentz2}
\end{eqnarray}

\begin{eqnarray}
\frac{\partial E_y}{\partial x}-\frac{\partial E_x}{\partial y}=-\frac{\partial B_z}{\partial t} \nonumber \\
\gamma\frac{\partial E_y}{\partial x'}-\frac{\partial E_x}{\partial y'}-\frac{\gamma v}{c^2}\frac{\partial E_y}{\partial t'}=
-\gamma\frac{\partial B_z}{\partial t'} +\gamma v\frac{\partial B_z}{\partial x'} \nonumber \\
\frac{\partial}{\partial x'}(\gamma E_y-\gamma vB_z)-\frac{\partial E_x}{\partial y'}=
-\frac{\partial }{\partial t'} (\gamma B_z - \frac{\gamma v}{c^2}E_y) \label{maxwelllorentz3}
\end{eqnarray}

Reescrevendo a lei de Gauss do campo el\'etrico (\ref{gausse}) e utilizando as derivadas parciais obtidas anteriomente, temos: 
\begin{eqnarray}
\nabla\cdot\vec{D}
=\frac{\partial D_x}{\partial x}+\frac{\partial D_y}{\partial y}+\frac{\partial D_z}{\partial z}=\rho \nonumber \\
\Rightarrow 
\gamma\frac{\partial D_x}{\partial x'}-\gamma\frac{v}{c^2}\frac{\partial D_x}{\partial t'}
+\frac{\partial D_y}{\partial y'}+\frac{\partial D_z}{\partial z'}=\rho \nonumber \\
\Rightarrow 
\gamma^2v\frac{\partial D_x}{\partial x'}=\gamma^2\frac{v^2}{c^2}\frac{\partial D_x}{\partial t'}
-\gamma v\frac{\partial D_y}{\partial y'}-\gamma v\frac{\partial D_z}{\partial z'}+\gamma v\rho
\end{eqnarray}

Al\'em disso, reescrevendo a lei de Ampere (\ref{ampere}) e utilizando as derivadas parciais obtidas anteriomente, temos: 
\begin{eqnarray}
\frac{\partial H_z}{\partial y}-\frac{\partial H_y}{\partial z}=J_x+\frac{\partial D_x}{\partial t} \nonumber \\
\frac{\partial H_z}{\partial y'}-\frac{\partial H_y}{\partial z'}
=J_x+\gamma\frac{\partial D_x}{\partial t'} -\gamma v\frac{\partial D_x}{\partial x'} \nonumber \\
\gamma\frac{\partial H_z}{\partial y'}-\gamma\frac{\partial H_y}{\partial z'}
=\gamma J_x+\gamma^2\frac{\partial D_x}{\partial t'} -\gamma^2 v\frac{\partial D_x}{\partial x'} \nonumber \\
\gamma\frac{\partial H_z}{\partial y'}-\gamma\frac{\partial H_y}{\partial z'}
=\gamma J_x - \gamma v\rho +\gamma^2\frac{\partial D_x}{\partial t'} -\gamma^2\frac{v^2}{c^2}\frac{\partial D_x}{\partial t'}
+\gamma v\frac{\partial D_y}{\partial y'}+\gamma v\frac{\partial D_z}{\partial z'}\nonumber \\
\frac{\partial }{\partial y'}(\gamma H_z-\gamma vD_y)-\frac{\partial }{\partial z'}(\gamma H_y+\gamma vD_z)
=\gamma(J_x - v\rho)+\gamma^2(1-\frac{v^2}{c^2})\frac{\partial D_x}{\partial t'} \label{amperelorentz1}
\end{eqnarray}

\begin{eqnarray}
\frac{\partial H_x}{\partial z}-\frac{\partial H_z}{\partial x}=J_y+\frac{\partial D_y}{\partial t} \nonumber \\
\frac{\partial H_x}{\partial z'}-\gamma\frac{\partial H_z}{\partial x'}+\frac{\gamma v}{c^2}\frac{\partial H_z}{\partial t'}=
J_y+\gamma\frac{\partial D_y}{\partial t'} -\gamma v\frac{\partial D_y}{\partial x'} \nonumber \\
\frac{\partial H_x}{\partial z'}
-\frac{\partial }{\partial x'}(\gamma H_z-\gamma vD_y)=
J_y+\frac{\partial }{\partial t'} (\gamma D_y-\frac{\gamma v}{c^2}H_z) \label{amperelorentz2}
\end{eqnarray}

\begin{eqnarray}
\frac{\partial H_y}{\partial x}-\frac{\partial H_x}{\partial y}=J_z+\frac{\partial D_z}{\partial t} \nonumber \\
\gamma\frac{\partial H_y}{\partial x'}-\frac{\partial H_x}{\partial y'}-\frac{\gamma v}{c^2}\frac{\partial H_y}{\partial t'}=
J_z+\gamma\frac{\partial D_z}{\partial t'} -\gamma v\frac{\partial D_z}{\partial x'} \nonumber \\
\frac{\partial}{\partial x'}(\gamma H_y+\gamma vD_z)-\frac{\partial H_x}{\partial y'}=
J_z+\frac{\partial }{\partial t'} (\gamma D_z - \frac{\gamma v}{c^2}H_y) \label{amperelorentz3}
\end{eqnarray}

Logo, para que as equa\c{c}\~oes (\ref{maxwelllorentz1}), (\ref{maxwelllorentz2}), (\ref{maxwelllorentz3}),
(\ref{amperelorentz1}), (\ref{amperelorentz2}) e (\ref{amperelorentz3}) 
sejam passadas para o referencial inercial $S'$, devemos fazer:
\begin{eqnarray}
E_x'=E_x \label{exlorentz} \\
E_y'=\gamma E_y-\gamma vB_z \label{eylorentz} \\
E_z'=\gamma E_z+\gamma vB_y \label{ezlorentz} \\
B_x'=B_x \label{bxlorentz} \\
B_y'=\gamma B_y+\frac{\gamma v}{c^2}E_z \label{bylorentz} \\
B_z'=\gamma B_z - \frac{\gamma v}{c^2}E_y \label{bzlorentz} \\
H_x'=H_x \label{hxlorentz} \\
H_y'=\gamma H_y+\gamma vD_z \label{hylorentz} \\
H_z'=\gamma H_z-\gamma vD_y \label{hzlorentz} \\
D_x'=D_x \label{dxlorentz} \\
D_y'=\gamma D_y-\frac{\gamma v}{c^2}H_z \label{dylorentz} \\
D_z'=\gamma D_z + \frac{\gamma v}{c^2}H_y \label{dzlorentz} \\
J_x'=\gamma J_x - \gamma v\rho \label{jxlorentz} \\
J_y'= J_y \label{jylorentz} \\
J_z'= J_z \label{jzlorentz} \\
\gamma^2(1-\frac{v^2}{c^2})=1 \label{gammalorentz} 
\end{eqnarray}

Observe que se multiplicarmos (\ref{exlorentz}), (\ref{eylorentz}) e (\ref{ezlorentz}) por $\varepsilon_0$ obteremos, 
respectivamente, (\ref{dxlorentz}), (\ref{dylorentz}) e (\ref{dzlorentz}), e 
se multiplicarmos (\ref{hxlorentz}), (\ref{hylorentz}) e (\ref{hzlorentz}) por $\mu_0$ obteremos, 
respectivamente, (\ref{bxlorentz}), (\ref{bylorentz}) e (\ref{bzlorentz}). O valor de $\gamma$ (denominado de
fator de Lorentz) pode ser obtido de (\ref{gammalorentz}):

\begin{equation}
\gamma^2(1-\frac{v^2}{c^2})=1 \Rightarrow  \gamma=\frac{1}{\sqrt{1-\frac{v^2}{c^2}}}
\label{fatordelorentz}
\end{equation}

Para a mec\^anica newtoniana tem-se que $v<<c$ e, portanto, $\gamma\approx 1$, 
fazendo com que a transforma\c{c}\~ao de Lorentz seja equivalente a transforma\c{c}\~ao galileana. 
Portanto, a transforma\c{c}\~ao de Lorentz satisfaz tanto \`a mec\^anica newtoniana, quanto \`a teoria eletromagn\'etica.
Para $v>c$ tem-se que $\gamma \rightarrow \infty$ e $v>c$ resultaria em $\gamma$ complexo.
Podemos observar tamb\'em a partir da transforma\c{c}\~ao de Lorentz, 
que o tempo n\~ao \'e absoluto como estabelecido na mec\^anica newtoniana,
e passa de maneira diferente nos diferentes referenciais inerciais.

\section{Teoria da relatividade especial}

Em 1905, Einstein prop\^os a teoria da relatividade especial com base em dois postulados:

1) As leis da f\'isica s\~ao equivalentes em todos os referenciais inerciais;

2) A velocidade da luz no v\'acuo \'e sempre igual a $c$.

\subsection{Intervalo relativ\'istico}

Dado um referencial $S$ formado pelos pontos $(ct,x,y,z)$, o deslocamento $\sqrt{(dx)^2+(dy)^2+(dz)^2}$ de um sinal luminoso
de $(0,0,0,0)$ para $(ct,x,y,z)$ \'e igual a $cdt=d\tau$, portanto:

\begin{equation} 
(d\tau)^2=(dx)^2+(dy)^2+(dz)^2.
\end{equation} 

O quadrado da dist\^ancia infinitesimal $ds^2$ em $S$ \'e definido como:

\begin{equation} 
(ds)^2=(d\tau)^2-(dx)^2-(dy)^2-(dz)^2.
\end{equation} 

Pelo segundo postulado, fica bem claro que
o quadrado da dist\^ancia percorrida pelo sinal luminoso em um outro referencial $S'$ que se move com velocidade $v$ constante 
em rela\c{c}\~ao a $S$ \'e:

\begin{equation} 
(d\tau')^2=(dx')^2+(dy')^2+(dz')^2.
\end{equation} 

Portanto, o quadrado da dist\^ancia infinitesimal $ds'^2$ em $S'$ \'e:

\begin{equation} 
(ds')^2=(d\tau')^2-(dx')^2-(dy')^2-(dz')^2.
\end{equation}

Logo, para um sinal luminoso, a forma de $ds^2$ \'e preservada, ou seja, $ds^2$ \'e invariante ($ds^2=ds'^2=0$). 
Para o movimento de uma part\'icula,
$ds^2$ continua invariante ($ds^2=ds'^2\ne0$)? 

Para responder essa pergunta, \'e preciso escrever $ds^2$ em termos das vari\'aveis $\tau'$, $x'$, $y'$ e $z'$.

\begin{eqnarray} 
\tau=\tau(\tau',x',y',z') \\ \nonumber
x=x(\tau',x',y',z') \\ \nonumber
y=y(\tau',x',y',z') \\ \nonumber
z=z(\tau',x',y',z')
\end{eqnarray} 

Pela regra da cadeia, tem-se:

\begin{eqnarray} 
d\tau=\frac{\partial\tau}{\partial\tau'}d\tau'
+\frac{\partial\tau}{\partial x'}dx'
+\frac{\partial\tau}{\partial y'}dy'
+\frac{\partial\tau}{\partial z'}dz' \\ \nonumber
d x=\frac{\partial x}{\partial\tau'}d\tau'
+\frac{\partial x}{\partial x'}dx'
+\frac{\partial x}{\partial y'}dy'
+\frac{\partial x}{\partial z'}dz' \\ \nonumber
d y=\frac{\partial y}{\partial\tau'}d\tau'
+\frac{\partial y}{\partial x'}dx'
+\frac{\partial y}{\partial y'}dy'
+\frac{\partial y}{\partial z'}dz' \\ \nonumber
d z=\frac{\partial z}{\partial\tau'}d\tau'
+\frac{\partial z}{\partial x'}dx'
+\frac{\partial z}{\partial y'}dy'
+\frac{\partial z}{\partial z'}dz'
\end{eqnarray} 

Calculando $(d\tau)^2$, $(dx)^2$, $(dy)^2$ e $(dz)^2$, tem-se:

\begin{eqnarray} 
(d\tau)^2=\frac{\partial\tau}{\partial\tau'}\frac{\partial\tau}{\partial\tau'}(d\tau')^2
+\frac{\partial\tau}{\partial x'}\frac{\partial\tau}{\partial x'}(dx')^2
+\frac{\partial\tau}{\partial y'}\frac{\partial\tau}{\partial y'}(dy')^2
+\frac{\partial\tau}{\partial z'}\frac{\partial\tau}{\partial z'}(dz')^2 \\ \nonumber
+2\frac{\partial\tau}{\partial\tau'}\frac{\partial\tau}{\partial x'}d\tau'dx'
+2\frac{\partial\tau}{\partial\tau'}\frac{\partial\tau}{\partial y'}d\tau'dy'
+2\frac{\partial\tau}{\partial\tau'}\frac{\partial\tau}{\partial z'}d\tau'dz' \\ \nonumber
+2\frac{\partial\tau}{\partial x'}\frac{\partial\tau}{\partial y'}dx'dy'
+2\frac{\partial\tau}{\partial x'}\frac{\partial\tau}{\partial z'}dx'dz' \\ \nonumber
+2\frac{\partial\tau}{\partial y'}\frac{\partial\tau}{\partial z'}dy'dz'
\end{eqnarray}

\begin{eqnarray} 
(d x)^2=\frac{\partial x}{\partial\tau'}\frac{\partial x}{\partial\tau'}(d\tau')^2
+\frac{\partial x}{\partial x'}\frac{\partial x}{\partial x'}(dx')^2
+\frac{\partial x}{\partial y'}\frac{\partial x}{\partial y'}(dy')^2
+\frac{\partial x}{\partial z'}\frac{\partial x}{\partial z'}(dz')^2 \\ \nonumber
+2\frac{\partial x}{\partial\tau'}\frac{\partial x}{\partial x'}d\tau'dx'
+2\frac{\partial x}{\partial\tau'}\frac{\partial x}{\partial y'}d\tau'dy'
+2\frac{\partial x}{\partial\tau'}\frac{\partial x}{\partial z'}d\tau'dz' \\ \nonumber
+2\frac{\partial x}{\partial x'}\frac{\partial x}{\partial y'}dx'dy'
+2\frac{\partial x}{\partial x'}\frac{\partial x}{\partial z'}dx'dz' \\ \nonumber
+2\frac{\partial x}{\partial y'}\frac{\partial x}{\partial z'}dy'dz'
\end{eqnarray} 

\begin{eqnarray} 
(d y)^2=\frac{\partial y}{\partial\tau'}\frac{\partial y}{\partial\tau'}(d\tau')^2
+\frac{\partial y}{\partial x'}\frac{\partial y}{\partial x'}(dx')^2
+\frac{\partial y}{\partial y'}\frac{\partial y}{\partial y'}(dy')^2
+\frac{\partial y}{\partial z'}\frac{\partial y}{\partial z'}(dz')^2 \\ \nonumber
+2\frac{\partial y}{\partial\tau'}\frac{\partial y}{\partial x'}d\tau'dx'
+2\frac{\partial y}{\partial\tau'}\frac{\partial y}{\partial y'}d\tau'dy'
+2\frac{\partial y}{\partial\tau'}\frac{\partial y}{\partial z'}d\tau'dz' \\ \nonumber
+2\frac{\partial y}{\partial x'}\frac{\partial y}{\partial y'}dx'dy'
+2\frac{\partial y}{\partial x'}\frac{\partial y}{\partial z'}dx'dz' \\ \nonumber
+2\frac{\partial y}{\partial y'}\frac{\partial y}{\partial z'}dy'dz'
\end{eqnarray} 

\begin{eqnarray} 
(d z)^2=\frac{\partial z}{\partial\tau'}\frac{\partial z}{\partial\tau'}(d\tau')^2
+\frac{\partial z}{\partial x'}\frac{\partial z}{\partial x'}(dx')^2
+\frac{\partial z}{\partial y'}\frac{\partial z}{\partial y'}(dy')^2
+\frac{\partial z}{\partial z'}\frac{\partial z}{\partial z'}(dz')^2 \\ \nonumber
+2\frac{\partial z}{\partial\tau'}\frac{\partial z}{\partial x'}d\tau'dx'
+2\frac{\partial z}{\partial\tau'}\frac{\partial z}{\partial y'}d\tau'dy'
+2\frac{\partial z}{\partial\tau'}\frac{\partial z}{\partial z'}d\tau'dz' \\ \nonumber
+2\frac{\partial z}{\partial x'}\frac{\partial z}{\partial y'}dx'dy'
+2\frac{\partial z}{\partial x'}\frac{\partial z}{\partial z'}dx'dz' \\ \nonumber
+2\frac{\partial z}{\partial y'}\frac{\partial z}{\partial z'}dy'dz'
\end{eqnarray} 

Portanto:

\begin{eqnarray} 
(ds)^2=\left(\frac{\partial\tau}{\partial\tau'}\frac{\partial\tau}{\partial\tau'}
+\frac{\partial x}{\partial\tau'}\frac{\partial x}{\partial\tau'}
+\frac{\partial y}{\partial\tau'}\frac{\partial y}{\partial\tau'}
+\frac{\partial z}{\partial\tau'}\frac{\partial z}{\partial\tau'}\right)(d\tau')^2 
+\left(\frac{\partial\tau}{\partial x'}\frac{\partial\tau}{\partial x'}
+\frac{\partial x}{\partial x'}\frac{\partial x}{\partial x'}
+\frac{\partial y}{\partial x'}\frac{\partial y}{\partial x'}
+\frac{\partial z}{\partial x'}\frac{\partial z}{\partial x'}\right)(dx')^2 + \\ \nonumber
+\left(\frac{\partial\tau}{\partial y'}\frac{\partial\tau}{\partial y'}
+\frac{\partial x}{\partial y'}\frac{\partial x}{\partial y'}
+\frac{\partial y}{\partial y'}\frac{\partial y}{\partial y'}
+\frac{\partial z}{\partial y'}\frac{\partial z}{\partial y'}\right)(dy')^2 
+\left(\frac{\partial\tau}{\partial z'}\frac{\partial\tau}{\partial z'}
+\frac{\partial x}{\partial z'}\frac{\partial x}{\partial z'}
+\frac{\partial y}{\partial z'}\frac{\partial y}{\partial z'}
+\frac{\partial z}{\partial z'}\frac{\partial z}{\partial z'}\right)(dz')^2 + \\ \nonumber
+2\left(\frac{\partial\tau}{\partial\tau'}\frac{\partial\tau}{\partial x'}
+\frac{\partial x}{\partial\tau'}\frac{\partial x}{\partial x'}
+\frac{\partial y}{\partial\tau'}\frac{\partial y}{\partial x'}
+\frac{\partial z}{\partial\tau'}\frac{\partial z}{\partial x'}\right)d\tau'dx' 
+2\left(\frac{\partial\tau}{\partial\tau'}\frac{\partial\tau}{\partial y'}
+\frac{\partial x}{\partial\tau'}\frac{\partial x}{\partial y'}
+\frac{\partial y}{\partial\tau'}\frac{\partial y}{\partial y'}
+\frac{\partial z}{\partial\tau'}\frac{\partial z}{\partial y'}\right)d\tau'dy' + \\ \nonumber
+2\left(\frac{\partial\tau}{\partial\tau'}\frac{\partial\tau}{\partial z'}
+\frac{\partial x}{\partial\tau'}\frac{\partial x}{\partial z'}
+\frac{\partial y}{\partial\tau'}\frac{\partial y}{\partial z'}
+\frac{\partial z}{\partial\tau'}\frac{\partial z}{\partial z'}\right)d\tau'dz' 
+2\left(\frac{\partial\tau}{\partial x'}\frac{\partial\tau}{\partial y'}
+\frac{\partial x}{\partial x'}\frac{\partial x}{\partial y'}
+\frac{\partial y}{\partial x'}\frac{\partial y}{\partial y'}
+\frac{\partial z}{\partial x'}\frac{\partial z}{\partial y'}\right)dx'dy' + \\ \nonumber
+2\left(\frac{\partial\tau}{\partial x'}\frac{\partial\tau}{\partial z'}
+\frac{\partial x}{\partial x'}\frac{\partial x}{\partial z'}
+\frac{\partial y}{\partial x'}\frac{\partial y}{\partial z'}
+\frac{\partial z}{\partial x'}\frac{\partial z}{\partial z'}\right)dx'dz' 
+2\left(\frac{\partial\tau}{\partial y'}\frac{\partial\tau}{\partial z'}
+\frac{\partial x}{\partial y'}\frac{\partial x}{\partial z'}
+\frac{\partial y}{\partial y'}\frac{\partial y}{\partial z'}
+\frac{\partial z}{\partial y'}\frac{\partial z}{\partial z'}\right)dy'dz'
\end{eqnarray} 

A magnitude da velocidade $v$ \'e a causa da varia\c{c}\~ao de $ds^2$, portanto, os coeficiente devem ser fun\c{c}\~oes
de $v$. A dire\c{c}\~ao da velocidade n\~ao causa nenhuma mudan\c{c}a em $ds^2$, devido a isotropia do espa\c{c}o-tempo 
(n\~ao existem dire\c{c}\~oes privilegiadas). Portanto, a equa\c{c}\~ao anterior pode ser reescrita como:

\begin{eqnarray} 
(ds)^2=f_{00}(v)(d\tau')^2 
+f_{11}(v)(dx')^2 
+f_{22}(v)(dy')^2 
+f_{33}(v)(dz')^2 
+f_{01}(v)d\tau'dx' + \\ \nonumber 
+f_{02}(v)d\tau'dy' 
+f_{03}(v)d\tau'dz' 
+f_{12}(v)dx'dy' 
+f_{13}(v)dx'dz' 
+f_{23}(v)dy'dz'
\end{eqnarray} 

Para obter uma express\~ao geral para $(ds)^2$, deve-se ter 
$(ds)^2=0\Rightarrow(ds')^2=0\Rightarrow(d\tau')^2=(dx')^2+(dy')^2+(dz')^2$. Portanto:

\begin{eqnarray} 
0=[f_{00}(v)+f_{11}(v)](dx')^2 
+[f_{00}(v)+f_{22}(v)](dy')^2 
+[f_{00}(v)+f_{33}(v)](dz')^2 
+f_{01}(v)d\tau'dx' + \\ \nonumber 
+f_{02}(v)d\tau'dy' 
+f_{03}(v)d\tau'dz' 
+f_{12}(v)dx'dy' 
+f_{13}(v)dx'dz' 
+f_{23}(v)dy'dz'
\end{eqnarray} 

Dessa forma, $f_{11}(v)=f_{22}(v)=f_{33}(v)=-f_{00}(v)=-g(v)$ e $f_{01}(v)=f_{02}(v)=f_{03}(v)=f_{12}(v)=f_{13}(v)=f_{23}(v)=0$.
Logo:

\begin{eqnarray} 
(d\tau)^2 -(dx)^2 -(dy)^2 -(dz)^2=g(v)\left[(d\tau')^2 -(dx')^2 -(dy')^2 -(dz')^2\right].
\end{eqnarray} 

A fun\c{c}\~ao $g(v)$ depende apenas da magnitude de $v$, portanto, a equa\c{c}\~ao anterior n\~ao se altera
quando o sentido de $v$ \'e invertido, o que equivale a permutar $S$ e $S'$ entre si. Dessa forma:

\begin{eqnarray} 
(d\tau')^2 -(dx')^2 -(dy')^2 -(dz')^2=g(v)\left[(d\tau)^2 -(dx)^2 -(dy)^2 -(dz)^2\right].
\end{eqnarray} 

Portanto, $g(v)=\pm 1$. A solu\c{c}\~ao $g(v)=- 1$ deve ser descartada, pois no caso de $v=0$, resultaria em $d\tau=-d\tau$,
$dx=-dx$, $dy=-dy$ e $dz=-dz$. Dessa
forma, o intervalo $ds=(d\tau)^2 -(dx)^2 -(dy)^2 -(dz)^2$ (chamado de \textit{intervalo relativ\'istico}) \'e invariante
nos diferentes referenciais inerciais. Logo:

\begin{eqnarray} 
ds'=ds.
\end{eqnarray} 

Dessa forma:

\begin{eqnarray} 
\int_{(ct'_1,x'_1,y'_1,z'_1)}^{(ct'_2,x'_2,y'_2,z'_2)}ds'=\int_{(ct_1,x_1,y_1,z_1)}^{(ct_2,x_2,y_2,z_2)}ds.
\end{eqnarray} 

Portanto:

\begin{eqnarray} 
(s'_2-s'_1)=(s_2-s_1) \Rightarrow \Delta s'=\Delta  s.
\end{eqnarray} 

No caso de $(ct'_1,x'_1,y'_1,z'_1)=(ct_1,x_1,y_1,z_1)=(0,0,0,0)$, tem-se:

\begin{eqnarray} 
s'= s.
\label{sls}
\end{eqnarray} 

Existem tr\^es casos poss\'iveis para $\Delta s$:

\begin{eqnarray} 
(\Delta s)^2 = 0 ~~~~~~~~~~\textrm{(Intervalo tipo nulo)} \\
(\Delta s)^2 > 0 ~~~~~~~~~~\textrm{(Intervalo tipo temporal)} \\
(\Delta s)^2 < 0 ~~~~~~~~~~\textrm{(Intervalo tipo espacial)} 
\end{eqnarray} 

Se um intervalo $\Delta s$ possui um determinado tipo num referencial $S$, devido \`a invari\^ancia de $ \Delta s$, ele possui
esse mesmo tipo em qualquer outro referencial $S'$.

Sobre o intervalo entre dois eventos, s\~ao pertinentes as seguintes observa\c{c}\~oes:

(1) Se os dois eventos s\~ao concernentes a um mesmo f\'oton, ent\~ao $(\Delta s)^2 = 0$, pois sempre
$c^2(\Delta t)^2=(\Delta x)^2+(\Delta y)^2+(\Delta z)^2$. 

(2) Se os dois eventos s\~ao concernentes a mesma part\'icula, existe um referencial $S'$ (fixo em rela\c{c}\~ao a part\'icula)
em que os eventos ocorrem na mesma posi\c{c}\~ao espacial, ou seja, $(\Delta x')^2+(\Delta y')^2+(\Delta z')^2=0$ e, portanto,
$(\Delta s')^2=c^2(\Delta t')^2>0$. Logo, nesse caso, o intervalo \'e tipo temporal. Pela invari\^ancia do intervalo, tem-se
$(\Delta s)^2=(c \Delta t)^2-(\Delta x)^2-(\Delta y)^2-(\Delta z)^2=c^2(\Delta t')^2>0$ 
$\Rightarrow\Delta t'=\frac{1}{c}\sqrt{c^2(\Delta t)^2-(\Delta x)^2-(\Delta y)^2-(\Delta z)^2}$. Al\'em disso,
$c^2>\frac{(\Delta x)^2+(\Delta y)^2+(\Delta z)^2}{(\Delta t)^2}$, ou seja, a part\'icula move-se com velocidade menor
do que $c$.

(2) Se os dois eventos s\~ao simult\^aneos em algum referencial $S'$, ou seja, $\Delta t'=0$, ent\~ao 
$(\Delta s')^2=-(\Delta x')^2-(\Delta y')^2-(\Delta z')^2<0$. Logo, nesse caso, o intervalo \'e tipo espacial,
e a dist\^ancia entre as posi\c{c}\~oes espaciais de ocorr\^encia dos eventos \'e 
$\sqrt{(\Delta x')^2+(\Delta y')^2+(\Delta z')^2}=\sqrt{(c\Delta t)^2-(\Delta x)^2-(\Delta y)^2-(\Delta z)^2}$.

\subsection{Transforma\c{c}\~ao de Lorentz}

Dado um referencial $S'$ que em rela\c{c}\~ao a $S$ move-se na dire\c{c}\~ao positiva de $x$ com velocidade $v$, de maneira que
$y'=y$ e $z'=z$.

\begin{eqnarray} 
x'=a_{11}x+a_{12}\tau \\
\tau'=a_{21}x+a_{22}\tau
\end{eqnarray} 

De \ref{sls}, tem-se:

\begin{eqnarray} 
\tau'^2-x'^2=\tau^2-x^2
\end{eqnarray} 

Portanto:

\begin{eqnarray} 
(a_{21}x+a_{22}\tau)^2-(a_{11}x+a_{12}\tau)^2=\tau^2-x^2
\end{eqnarray} 

\begin{eqnarray} 
(a_{22}^2-a_{12}^2)\tau^2-(a_{11}^2-a_{21}^2)x^2+2(a_{21}a_{22}-a_{11}a_{12})\tau x=\tau^2-x^2
\end{eqnarray} 

\begin{eqnarray} 
a_{22}^2-a_{12}^2=1 \\
a_{11}^2-a_{21}^2=1 \\
a_{21}a_{22}=a_{11}a_{12}
\end{eqnarray} 

\begin{eqnarray} 
1-\left(\frac{a_{12}}{a_{22}}\right)^2=\frac{1}{a_{22}^2} \label{1eq} \\
1-\left(\frac{a_{21}}{a_{11}}\right)^2=\frac{1}{a_{11}^2} \label{2eq} \\
\frac{a_{12}}{a_{22}}=\frac{a_{21}}{a_{11}} \label{3eq}
\end{eqnarray} 

Substituindo (\ref{3eq}) em (\ref{1eq}) e comparando o resultado com (\ref{1eq}), tem-se:

\begin{eqnarray} 
a_{22}^2=a_{11}^2 \Rightarrow a_{22}=\pm a_{11}
\end{eqnarray} 

Se $v=0$, ent\~ao $x=x'$ e $\tau=\tau'$ e, portanto, $a_{22}\ne -a_{11}$. Dessa forma:

\begin{eqnarray} 
a_{22}=a_{11}=\gamma.
\label{gamma1}
\end{eqnarray} 

Uma part\'icula que se mant\'em fixa em $x=0$ com o passar do tempo $t$ em um referencial $S$, \'e vista por um
ponto fixo no referencial $S'$ que se move com velocidade $v$ para a direita, como se estivesse
em movimento para a esquerda com velocidade $-v=x'/t'$. Al\'em disso, fazendo $x=0$, tem-se:

\begin{eqnarray} 
\frac{a_{12}}{a_{22}}=\frac{x'}{\tau'}=\frac{x'}{t'}\frac{1}{c}=\frac{-v}{c}.
\label{vc1}
\end{eqnarray} 

Substituindo (\ref{gamma1}) e (\ref{vc1}) em (\ref{1eq}), tem-se:

\begin{equation} 
\gamma = \frac{1}{ \sqrt{1 - { \frac{v^2}{c^2}}}}
\end{equation} 

Se a part\'icula est\'a em repouso ($v=0$), ent\~ao $\gamma = 1$.

Como a express\~ao $( 1 + a)^{n}\approx (1 + n  a)$ \'e v\'alida para $a<<1$,
para part\'iculas em baixas velocidades ($v<<c$) tem-se:

\begin{equation} 
\gamma = \frac{1}{ \sqrt{1 - { \frac{v^2}{c^2}}}}=\left ( 1 - { \frac{v^2}{c^2}} \right)^{-1/2}
\approx 1 +\frac{1}{2}  \frac{v^2}{c^2}
\label{fatordelorentzaproximado}
\end{equation} 

A transforma\c{c}\~ao de Lorentz \'e obtida como:

\begin{eqnarray} 
x'=\gamma x-\frac{v\gamma}{c}\tau \\
\tau'= \gamma \tau-\frac{v\gamma}{c}x
\label{tranformacaodeeinsteinlorentz}
\end{eqnarray} 

\begin{eqnarray} 
x'=\gamma( x- vt) \\
t'= \gamma( t-\frac{v}{c^2}x)
\end{eqnarray} 

\subsection{Simultaneidade}

O gr\'afico da posi\c{c}\~ao da part\'icula em fun\c{c}\~ao do tempo \'e chamado de \textit{linha de mundo}, e
cada ponto desse gr\'afico \'e um \textit{ponto de universo}.
Na figura \ref{fig.simultaneidade1}, s\~ao mostradas as linhas de mundo para tr\^es part\'iculas em repouso em rela\c{c}\~ao
ao referencial $S$, as quais est\~ao localizadas
nos pontos $A$, $B$ e $C$ (linhas verticais). Al\'em disso, s\~ao mostradas as linhas de mundo de um sinal de r\'adio,
o qual \'e enviado de $B$ em $t=0$. Como mostrado na figura, os eventos "chegada do sinal em $A$" e 
"chegada do sinal em $B$" s\~ao simult\^aneos, pois, ocorrem em tempos iguais $t_A=t_B$.

\begin{figure}[!h]
\begin{center}
\includegraphics[scale=0.3, bb = 250 80 400 450]{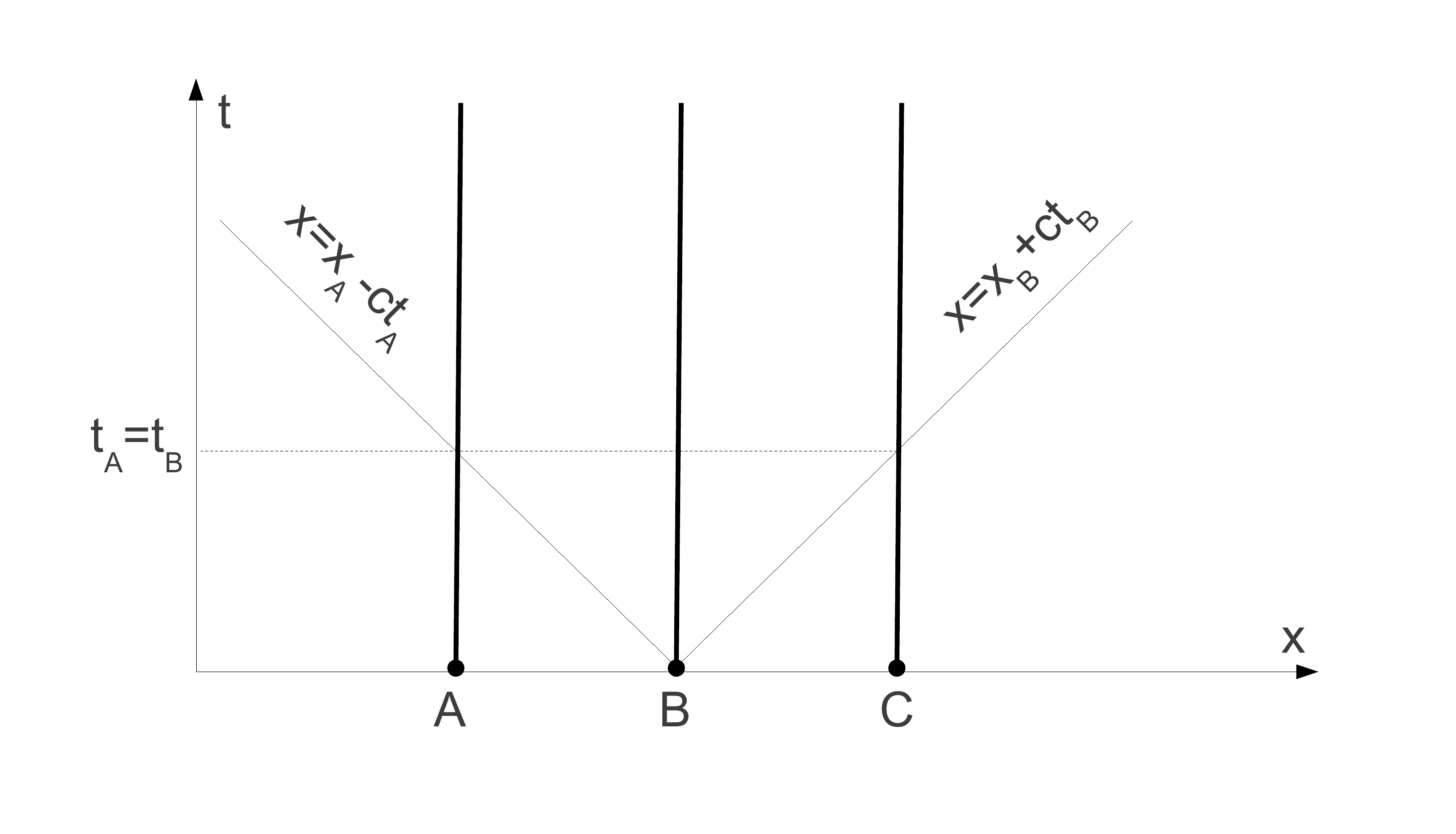}
\end{center}
\caption{Os eventos "chegada do sinal em $A$" e 
"chegada do sinal em $B$" s\~ao simult\^aneos, pois, ocorrem em tempos iguais $t_A=t_B$ no referencial $S$.}
\label{fig.simultaneidade1}
\end{figure}

\begin{figure}[!h]
\begin{center}
\includegraphics[scale=0.3, bb = 250 100 400 450]{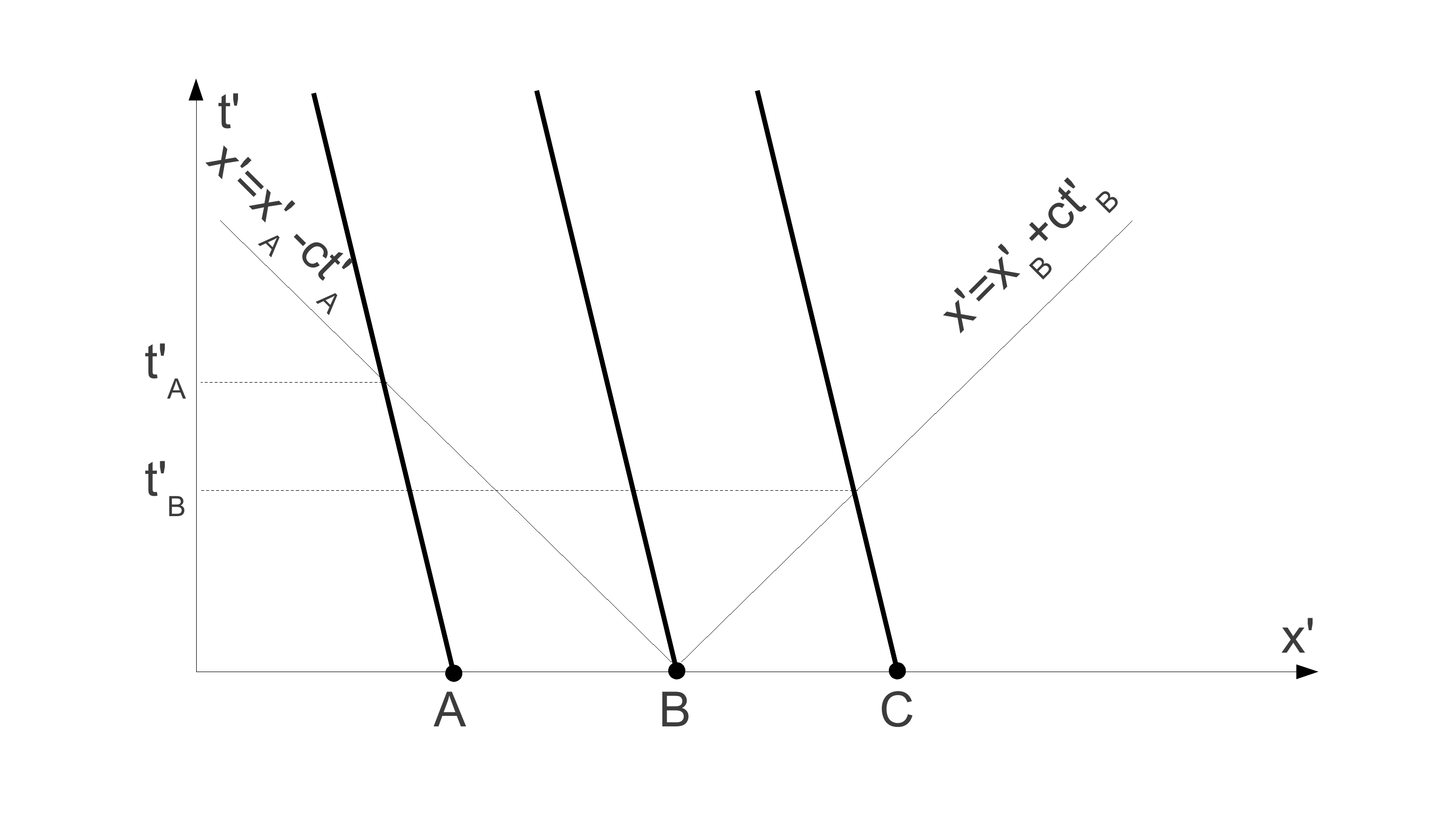}
\end{center}
\caption{Os eventos "chegada do sinal em $A$" e "chegada do sinal em $B$" n\~ao s\~ao simult\^aneos, 
pois, ocorrem em tempos diferentes $t_A\ne t_B$  no referencial $S'$.}
\label{fig.simultaneidade2}
\end{figure}

Essas tr\^es part\'iculas, em um referencial $S'$ que se desloca para a direita com rela\c{c}\~ao ao referencial $S$,
est\~ao em repouso em rela\c{c}\~ao a $S$, mas est\~ao deslocando para a esquerda em rela\c{c}\~ao a $S'$, a medida
que o tempo $t'$ passa,
como mostrado na figura \ref{fig.simultaneidade2}.
Os eventos "chegada do sinal em $A$" e "chegada do sinal em $B$" n\~ao s\~ao simult\^aneos, 
pois, ocorrem em tempos diferentes $t_A\ne t_B$.

Portanto, ao afirmar que dois eventos s\~ao simult\^aneos, \'e necess\'ario especificar em qual referencial essa 
afirma\c{c}\~ao \'e verdadeira. 

A exist\^encia de um referencial em que os eventos s\~ao simult\^aneos, permite criar um 
\textit{m\'etodo para sincronizar dois rel\'ogios}. 
Se dois rel\'ogios s\~ao colocados nos pontos $A$ e $C$, e esses rel\'ogios s\~ao ligados no instante em que
o sinal vindo de $B$ os atinge, eles estar\~ao sincronizados no referencial $S$.

\subsection{Velocidade m\'axima das intera\c{c}\~oes e causalidade}

A trajet\'oria descrita por uma part\'icula que se desloca em MRU (movimento retil\'ineo uniforme) ao longo do eixo $x$, 
passando pelo ponto de universo $O\equiv(t,x)=(0,0)$, \'e uma reta (como a reta tracejada mostrada na figura 
\ref{fig.conedeluz}), cuja tangente do \^angulo formado com o eixo $t$ \'e a velocidade da part\'icula. 
As equa\c{c}\~oes $x=ct$ e $x=-ct$ descrevem as trajet\'orias dos f\'otons que se movem, respectivamente, nos sentidos
$+x$ e $-x$. Se fossem mostradas as outras coordenadas a superf\'icie
$x^2+y^2+z^2=ct$ corresponderia ao \textit{cone de luz}.

Os intervalos entre dois eventos nas regi\~oes azul e vermelha s\~ao do tipo temporal, portanto, 
n\~ao h\'a eventos simult\^aneos nelas. A regi\~ao
azul $(t>0)$ \'e chamada de \textit{futuro absoluto relativamente a O}, pois nela todos os eventos ocorrem "depois" de $O$.
A regi\~ao vermelha $(t<0)$ \'e chamada de \textit{passado absoluto relativamente a O},
pois nela todos os eventos ocorrem "antes" de $O$.
Dessa forma, dois eventos contidos nas regi\~oes azul ou vermelha s\~ao \textit{causalmente relacionados}, e
as intera\c{c}\~oes propagam numa velocidade menor do que $c$.

\begin{figure}[!h]
\begin{center}
\includegraphics[scale=0.3, bb = 800 140 0 560]{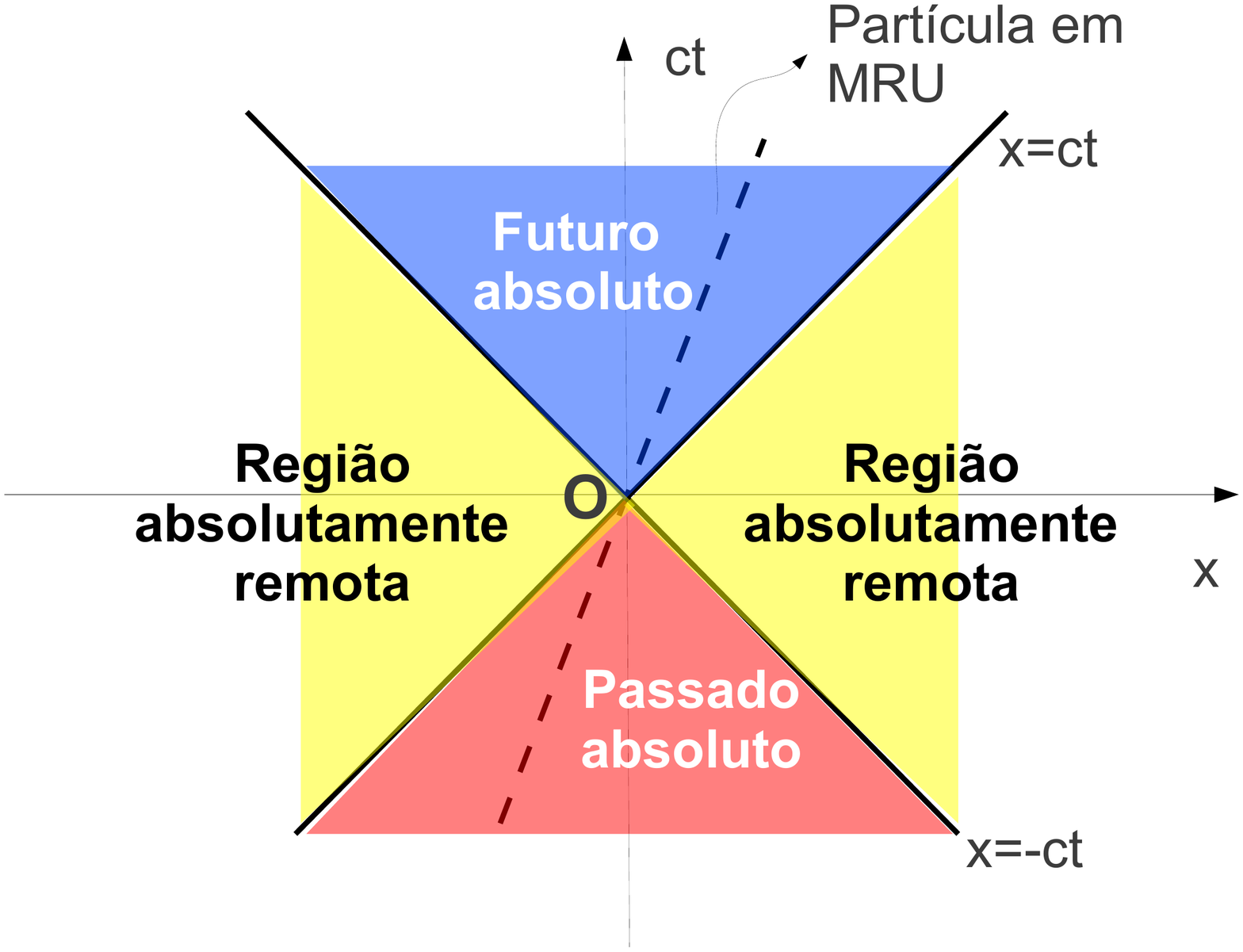}
\end{center}
\caption{Cone de luz.}
\label{fig.conedeluz}
\end{figure}

O intervalo entre $O$ e um ponto $P$ qualquer da regi\~ao amarela \'e do tipo espacial, portanto,
os eventos $O$ e $P$ sempre representam diferentes pontos espaciais. Por isso,
a regi\~ao amarela \'e chamada de \textit{regi\~ao absolutamente remota relativamente a O}.
O evento $P$ pode ser simult\^aneo a $O$, estar no futuro de $O$, ou no passado de $O$, dependendo do
referencial. Dessa forma, nessa regi\~ao os eventos n\~ao s\~ao \textit{causalmente relacionados}.

\subsection{Dilata\c{c}\~ao temporal}

Um rel\'ogio move-se com velocidade $v$ em um referencial $S$, percorrendo uma dist\^ancia 
$\sqrt{(\Delta x)^2+(\Delta y)^2+(\Delta z)^2}$ num tempo $\Delta t$. Em um referencial $S'$
que se move com rela\c{c}\~ao a $S$, mas fixo em rela\c{c}\~ao ao rel\'ogio, tem-se
$\Delta x'=\Delta y'=\Delta z'=0$, e $\Delta t'$ \'e chamado de tempo pr\'oprio. 
Devido a invari\^ancia do intervalo relativ\'istico tem-se:

\begin{eqnarray} 
(\Delta s')^2=(\Delta s)^2 \\
(c\Delta t')^2-(\Delta x')^2-(\Delta y')^2-(\Delta z')^2=(c\Delta t)^2-(\Delta x)^2-(\Delta y)^2-(\Delta z)^2 \\
(c\Delta t')^2=(c\Delta t)^2-(\Delta x)^2-(\Delta y)^2-(\Delta z)^2 \\
(\Delta t')^2=(\Delta t)^2\left[1-\frac{1}{c^2}\frac{(\Delta x)^2-(\Delta y)^2-(\Delta z)^2}{(\Delta t)^2}\right] \\
(\Delta t')^2=(\Delta t)^2\left[1-\frac{v^2}{c^2}\right] \Rightarrow
\Delta t=\gamma\Delta t'
\end{eqnarray} 

Dessa forma, no referencial em que o rel\'ogio est\'a em movimento, o tempo parecer\'a dilatado. 
Contudo, \'e um erro querer comparar o tempo de rel\'ogios espacialmente afastados, pois,
a informa\c{c}\~ao do tempo de um rel\'ogio teria de chegar instantaneamente no outro rel\'ogio, o que n\~ao \'e poss\'ivel.
Portanto, para observar a dilata\c{c}\~ao temporal de maneira correta, \'e preciso dispor de um conjunto de rel\'ogios
sincronizados, espacialmente afastados,
os quais podem ser obtidos atrav\'es do \textit{m\'etodo para sincronizar dois rel\'ogios}
descrito anteriomente, em rela\c{c}\~ao ao qual o rel\'ogio em movimento \'e comparado.

Experimentos com rel\'ogios at\^omicos de elevada precis\~ao em avi\~oes corroboraram a dilata\c{c}\~ao do tempo
prevista pela relatividade especial.

\subsubsection{Paradoxo dos g\^emeos}

Se dois irm\~aos g\^emeos ligarem os cron\^ometros de seus rel\'ogios no momento em que um deles parte em uma viagem espacial
em uma nave que em $\Delta t'=9$ anos se afasta de $7,2$ anos-luz da Terra at\'e uma esta\c{c}\~ao espacial, 
para o irm\~ao que ficou na Terra, ter\~ao
passado $\Delta t=15$ anos, e ser\'a $12$ anos mais velho que o irm\~ao, quando ele retornar
(isso \'e conhecido como \textit{paradoxo dos g\^emeos}). Todavia, 
o g\^emeo-astronauta poderia contestar que ele \'e quem seria o mais velho, pois,
no referencial $S'$ do seu rel\'ogio, o g\^emeo-n\~ao-astronauta \'e quem estaria em movimento.
Por\'em, o g\^emeo-n\~ao-astronauta \'e quem \'e um observador inercial, e pode contar com um
conjunto de rel\'ogios fixos e sincronizados distribuidos ao longo do espa\c{c}o. Ao contr\'ario do g\^emeo-astronauta
que acelera ao sair da Terra, ao inverter o sentido do movimento e ao pousar na Terra, alterando constantemente o seu 
referencial. Igualmente, 
caso o g\^emeo-astronauta estivesse fazendo uma viagem espacial em uma trajet\'oria em \'orbita em torno da Terra, 
os referenciais seriam inerciais apenas sob o ponto de vista do g\^emeo-n\~ao-astronauta.

\subsubsection{Decaimento de m\'uons}

Os r\'aios c\'osmicos provenientes do espa\c{c}o sideral d\~ao origem a m\'uons na atmosfera terrestre, os quais se deslocam com
velocidades m\'edias $v=0,992c$, indo em dire\c{c}\~ao a superf\'icie.
A vida m\'edia de um m\'uon em repouso \'e $T=2,2\times 10^-6~s$. Em 1963, em um experimento, foram detectados $550$ m\'uons
no alto do Monte Washington, numa altitude $h=1907~m$, 
e no mesmo per\'iodo foram detectados $397$ m\'uons ao n\'ivel do mar. Utilizando a equa\c{c}\~ao de decaimento, dado
que $N_0$ \'e o n\'umero de m\'uons no instante $t_0=0$, o n\'umero $N$ de m\'uons decorrido um tempo $t$ \'e:

\begin{equation}
N=N_0~exp(-\frac{t}{T})=N_0~exp(-\frac{h/v}{T})\approx 30 ~\textrm{m\'uons}.
\end{equation}
Nota-se que esse resultado n\~ao est\'a de acordo com o n\'umero de part\'iculas que foram detectadas. Todavia, 
pela relatividade especial, deve ocorrer a dilata\c{c}\~ao do tempo para uma part\'icula em movimento, portanto,
para a part\'icula o tempo decorrido $t'$ foi menor, dado por:

\begin{equation}
t'=\frac{t}{\gamma}=\frac{t}{\frac{1}{\sqrt{1-\frac{v^2}{c^2}}}}=\frac{t}{7,9216}.
\end{equation}
Portanto:

\begin{equation}
N=N_0~exp(-\frac{t'}{T})=N_0~exp(-\frac{h/v}{7,9216\times T})\approx 381 ~\textrm{m\'uons}.
\end{equation}

Considerando que esses valores s\~ao aproximados, v\^e-se que a dilata\c{c}\~ao temporal consegue explicar 
o fato de os m\'uos mesmo sendo t\~ao inst\'aveis conseguirem atingir a superf\'icie da Terra.

\section{Transforma\c{c}\~ao de velocidades}

Considere um referencial $S$, e um segundo referencial $S'$ que se move para a direita ao longo do eixo 
$x$ com velocidade $v$ constante em rela\c{c}\~ao ao referencial $S$. Um evento \'e representado respectivamente por 
$(ct,x,y,z)$ e $(ct',x',y',z')$ nos referenciais $S$ e $S'$. As coordenadas se 
relacionam pela transforma\c{c}\~ao de Lorentz (\ref{lorentz}). 
Ao derivar a equa\c{c}\~ao (\ref{lorentz}) com rela\c{c}\~ao a $t$, obt\'em-se:

\begin{eqnarray}
\frac{d}{dt}x'=\frac{d}{dt}\left[\gamma(x-vt) \nonumber \right]=\gamma\left(\frac{dx}{dt}-v\right) \nonumber \\
\frac{d}{dt}y'=\frac{d}{dt}y \nonumber \\
\frac{d}{dt}z'=\frac{d}{dt}z \nonumber \\
\frac{d}{dt}t'=\frac{d}{dt}\left[\gamma(t-vx/c^2)  \nonumber \right]=\gamma\left(1-\frac{v}{c^2}\frac{dx}{dt}\right)
\label{dlorentz}
\end{eqnarray}

Se $\vec{u}=(u_x,u_y,u_z)$ e $\vec{u'}=(u'_x,u'_y,u'_z)$ s\~ao os vetores velocidade de uma part\'icula respectivamente 
nos referenciais $S$ e $S'$, ent\~ao:

\begin{eqnarray}
u'_x=\frac{dx'}{dt'}=\frac{dx'}{dt}\frac{dt}{dt'}=\frac{\frac{dx'}{dt}}{\frac{dt'}{dt}}
=\frac{\gamma\left(\frac{dx}{dt}-v\right)}{\gamma\left(1-\frac{v}{c^2}\frac{dx}{dt}\right)}
=\frac{u_x-v}{1-\frac{v}{c^2}u_x} \\
u'_y=\frac{dy'}{dt'}=\frac{dy'}{dt}\frac{dt}{dt'}=\frac{dy}{dt}\frac{dt}{dt'}=\frac{dy}{dt}\frac{dt}{dt'} 
=\frac{u_y}{\gamma\left(1-\frac{v}{c^2}u_x\right)} \\
u'_z=\frac{dz'}{dt'}=\frac{dz'}{dt}\frac{dt}{dt'}=\frac{dz}{dt}\frac{dt}{dt'}=\frac{dz}{dt}\frac{dt}{dt'} 
=\frac{u_z}{\gamma\left(1-\frac{v}{c^2}u_x\right)} \\
\frac{d}{dt}t'=\frac{d}{dt}\left[\gamma(t-vx/c^2)  \right]=\gamma\left(1-\frac{v}{c^2}\frac{dx}{dt}\right)
=\gamma\left(1-\frac{v}{c^2}u_x\right)
\label{vlorentz}
\end{eqnarray}

No limite cl\'assico ($c\rightarrow \infty$), tem-se:

\begin{equation}
u'_x=u_x-v \textrm{,}~~u'_y=u_y ~~\textrm{e}~~u'_z=u_z.
\end{equation}

Fazendo $u_x=\frac{\Delta x}{\Delta t}$, $u_y=\frac{\Delta y}{\Delta t}$, $u_z=\frac{\Delta z}{\Delta t}$, 
$u'_x=\frac{\Delta x'}{\Delta t'}$, $u'_y=\frac{\Delta y'}{\Delta t'}$ e $u'_z=\frac{\Delta z'}{\Delta t'}$, temos:

\begin{eqnarray}
\frac{\Delta x'}{\Delta t'}=\frac{\frac{\Delta x}{\Delta t}-v}{1-\frac{v}{c^2}\frac{\Delta x}{\Delta t}} 
~~~~~~~~~~~~~~~~~~~~~~~~~\Delta x'=\gamma(\Delta x-v\Delta t) 
\label{eq.deltax} \\
\frac{\Delta y'}{\Delta t'}=\frac{\frac{\Delta y}{\Delta t}}{\gamma\left(1-\frac{v}{c^2}\frac{\Delta x}{\Delta t}\right)} 
\Rightarrow~~~~~~~~~~~~~~~~~~~~~~~~~~~~~\Delta y'=\Delta y \\
\frac{\Delta z'}{\Delta t'}=\frac{\frac{\Delta z}{\Delta t}}{\gamma\left(1-\frac{v}{c^2}\frac{\Delta x}{\Delta t}\right)} 
~~~~~~~~~~~~~~~~~~~~~~~~~~~~~~~~~~\Delta z'=\Delta z \\
\frac{\Delta t'}{\Delta t}=\gamma\left(1-\frac{v}{c^2}\frac{\Delta x}{\Delta t}\right)
~~~~~~~~~~~~~~~~\Delta t'=\gamma(\Delta t-\frac{v}{c^2}\Delta x)
\label{v1lorentz}
\end{eqnarray} 

\subsection{Contra\c{c}\~ao das dist\^ancias}

Seja uma r\'egua localizada ao longo do eixo $x$.
Para dois eventos no referencial em repouso $S$, um deles em $(ct_i,x_i,y_i,z_i)$, localizado na extremidade inicial da r\'egua, 
e o outro em $(ct_f,x_f,y_f,z_f)$, localizado na extremidade final da r\'egua,  
os quais podem ser representados respectivamente em $S'$ como $(ct'_i,x'_i,y'_i,z'_i)$ e $(ct'_f,x'_f,y'_f,z'_f)$, com $S'$
sendo um referencial que se move para a direita ao longo do eixo $x$ com velocidade $v$ constante em rela\c{c}\~ao 
ao referencial $S$.

Obviamente, o tamanho da r\'egua em $S$ \'e $L_0=x_f-x_i$. O tamanho da r\'egua $L=x'_f-x'_i$ \'e obtido no referencial $S'$
quando os eventos inicial e final no referencial $S'$ forem simult\^aneos $(t'_f=t'_i\Rightarrow \Delta t'=t'_f-t'_i=0)$.
Permutar $S$ e $S'$ equivale a trocar $v$ por $-v$, portanto, \'e poss\'ivel reescrever (\ref{eq.deltax})
como

\begin{equation}
\Delta x = \gamma(\Delta x' + v\Delta t'). 
\end{equation} 

Dessa forma, o comprimento $L$ da r\'egua no referencial $S'$ \'e:

\begin{equation}
\Delta x = \gamma(\Delta x' + v\Delta t') \Rightarrow  L_0 = \gamma L \Rightarrow  L = \frac{L_0}{\gamma}.
\end{equation} 

Portanto, os comprimentos s\~ao contra\'idos na dire\c{c}\~ao do movimento.

\subsection{Aberra\c{c}\~ao relativ\'istica}

O efeito de \textit{aberra\c{c}\~ao da luz} ilustrado na figura \ref{fig.aberracao2} \'e explicado pela
relatividade especial. Considerando uma part\'icula que se move sobre o plano $xy$, tem-se:

\begin{equation}
tan\theta'=\frac{sen\theta'}{cos\theta'}=\frac{u'~sen\theta'}{u'~cos\theta'}=\frac{u'_y}{u'_x}
=\frac{\frac{u_y}{\gamma\left(1-\frac{v}{c^2}u_x\right)}}{\frac{\gamma\left(u_x-v\right)}{\gamma\left(1-\frac{v}{c^2}u_x\right)}}
=\frac{u_y}{\gamma\left(u_x-v\right)}=\frac{u~sen\theta}{\gamma\left(u~cos\theta-v\right)}.
\end{equation} 

Para um f\'oton $u=u'=c$. Portanto:

\begin{eqnarray}
\frac{sen\theta'}{cos\theta'}=\frac{sen\theta}{\gamma\left(cos\theta-\frac{v}{c}\right)} \nonumber \\
{sen\theta'}{\gamma\left(cos\theta-\frac{v}{c}\right)}={sen\theta}{cos\theta'} \nonumber \\
\gamma {sen\theta'} cos\theta-{sen\theta}{cos\theta'}=\frac{v}{c}{sen\theta'}
\end{eqnarray} 

Como $v<<c$:

\begin{eqnarray}
\left(1-\frac{v^2}{2c^2}\right) {sen\theta'} cos\theta-{sen\theta}{cos\theta'}=\frac{v}{c}{sen\theta'} \nonumber \\
sen(\theta'-\theta)-\frac{v^2}{2c^2} {sen\theta'} cos\theta=\frac{v}{c}{sen\theta'}
\end{eqnarray} 

Fazendo $\Delta \theta=\theta'-\theta$, e $sen \Delta \theta \approx \Delta \theta$ (pois $\Delta \theta \approx 0$):

\begin{equation}
\Delta\theta=\frac{v}{c}{sen\theta'}+\frac{v^2}{2c^2} {sen\theta'} cos\theta.
\end{equation}

Al\'em disso, se $\frac{v^2}{c^2}\approx 0$:

\begin{equation}
\Delta\theta=\frac{v}{c}{sen\theta'}.
\end{equation}

Essa express\~ao coincide com a que foi proposta por James Bradley em 1829 para a \textit{aberra\c{c}\~ao estelar},
a qual foi utilizada no c\'alculo de $c$.

\subsection{Efeito Doppler}

Dado um referencial $S$ de cuja origem uma fonte envia pulsos de radar: o primeiro pulso \'e enviado em $t=0$ com
o observador em $x=x_0$ e o $(n+1)$-\'esimo pulso \'e enviado quando $t=n\tau$, pois a frequ\^encia da fonte \'e
$f=1/\tau$. O observador move-se com velocidade $v$ e est\'a em repouso em rela\c{c}\~ao ao referencial $S'$.
Todas essas informa\c{c}\~oes s\~ao mostradas no gr\'afico da figura \ref{fig.doppler}. Dessa forma, tem-se:

\begin{eqnarray}
x_1=ct_1=x_0+vt_1 \\
x_2=c(t_2-n\tau)=x_0+vt_2
\end{eqnarray}

Portanto:

\begin{eqnarray}
t_2-t_1=\frac{nc\tau}{c-v} \\
x_2-x_1=c(t_2-t_1)-nc\tau \Rightarrow x_2-x_1=\frac{ncv\tau}{c-v}
\end{eqnarray}

\begin{figure}[!h]
\begin{center}
\includegraphics[scale=0.3, bb = 100 100 600 550]{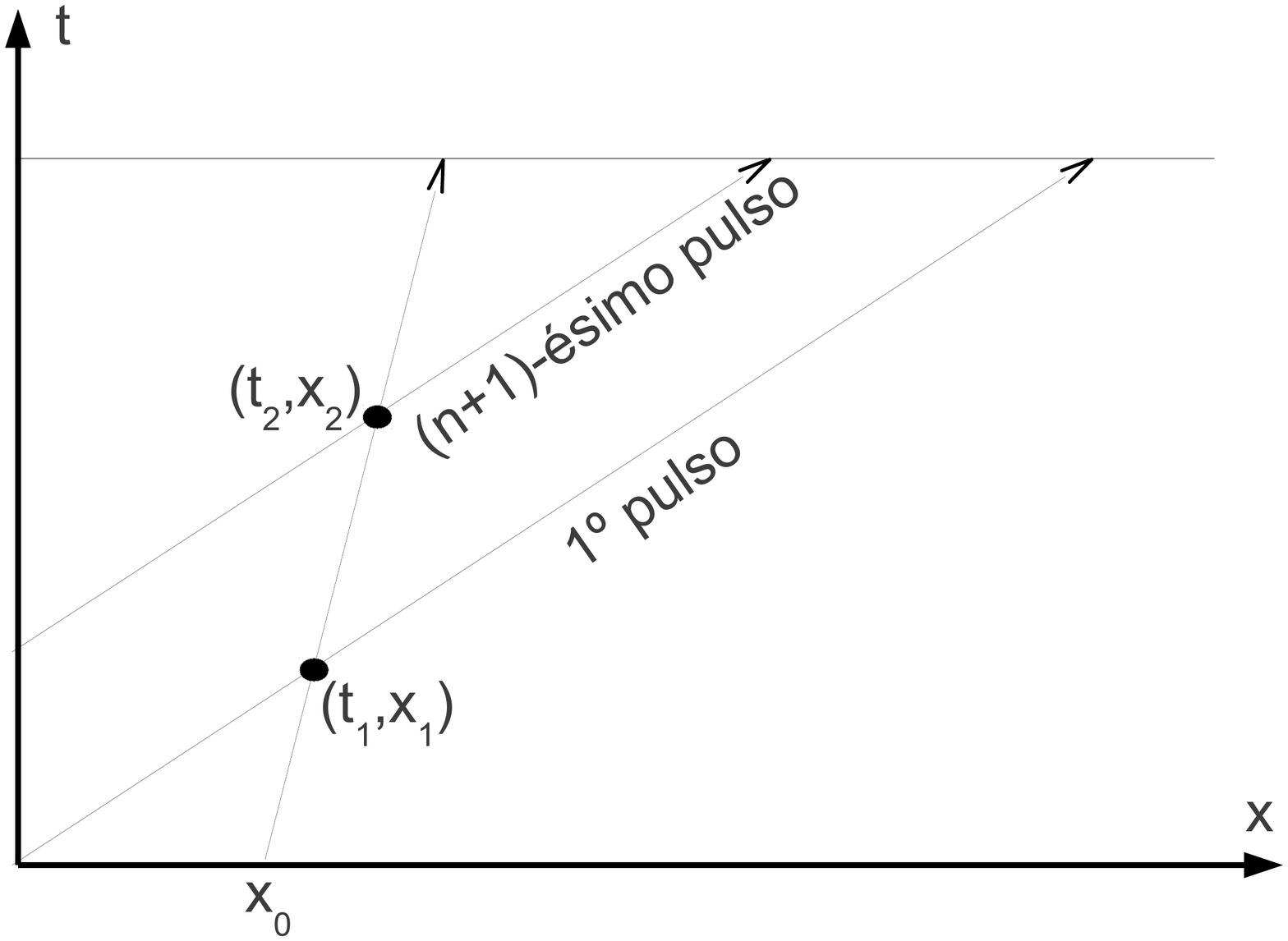}
\end{center}
\caption{Efeito Doppler}
\label{fig.doppler}
\end{figure}

Pela transforma\c{c}\~ao de Lorentz, tem-se:

\begin{eqnarray}
t'_2-t'_1=\gamma \left[(t_2-t_1)-v(x_2-x_1)/c^2\right]
=\gamma \left[\frac{nc\tau}{c-v}-\frac{v}{c^2}\frac{ncv\tau}{c-v}\right]
=\gamma \frac{nc\tau}{c-v} \left[1-\frac{v}{c^2}\right]
\end{eqnarray}

Assim, o per\'iodo aparente do sinal recebido pelo observador \'e:

\begin{eqnarray}
\tau'=\frac{t'_2-t'_1}{n}=\gamma \frac{c\tau}{c-v} \left[1-\frac{v}{c^2}\right]=\gamma(1+\beta)\tau
\label{periodoaparente}
\end{eqnarray} onde $\beta=v/c$. Como $\gamma=(1-\beta)^{-1/2}$, tem-se:

\begin{eqnarray}
\tau'=\left(\frac{1+\beta}{1-\beta}\right)^{1/2}\tau
\end{eqnarray}

Em termos da frequ\^encia, tem-se:

\begin{eqnarray}
f'=\left(\frac{1-\beta}{1+\beta}\right)^{1/2}f
\label{eq.doppler}
\end{eqnarray}

Em termos do comprimento de onda, tem-se:

\begin{eqnarray}
\lambda'=\frac{c}{f'}=\left(\frac{1+\beta}{1-\beta}\right)^{1/2}\frac{c}{f}=\left(\frac{1+\beta}{1-\beta}\right)^{1/2}\lambda
\label{eq.doppler1}
\end{eqnarray}

Fazendo $\gamma=1$ (caso cl\'assico) em (\ref{periodoaparente}), e multiplicando por $c$, tem-se:

\begin{eqnarray}
\lambda'=(1+\beta)\lambda
\label{eq.dopplerc}
\end{eqnarray}

\subsubsection{Deslocamento para o vermelho}

Da equa\c{c}\~ao \ref{eq.doppler1}, tem-se:

\begin{eqnarray}
\beta=\frac{(\lambda'/\lambda)^2-1}{(\lambda'/\lambda)^2+1}
\label{velocidadeafastamento}
\end{eqnarray}

Em 1929, Hubble observou que quanto mais distante est\'a uma gal\'axia maior \'e o seu deslocamento para o vermelho
(transla\c{c}\~ao das linhas espectrais no sentido em que ocorre o aumento dos comprimentos de onda), e interpretou
isso como uma evid\^encia de que o universo est\'a em expans\~ao. Por exemplo, considerando as linhas espectrais do 
hidrog\^enio, se a linha espectral com $\lambda'=656~nm$ \'e deslocada para $\lambda=1458~nm$, ent\~ao a velocidade
de afastamento da gal\'axia, calculada a partir de \ref{velocidadeafastamento}, \'e $v=0,664c$.

\subsubsection{Experimento de Ives-Stilwell}

Em 1938, Ives e Stilwell realizaram um experimento em que comprovaram a dilata\c{c}\~ao temporal \cite{ives}. 
Descargas em g\'as hidrog\^enio produzem \'ions $H_2^{+}$ e $H_3^{+}$, os quais atravessam um campo el\'etrico 
com cerca de 10 keV, e s\~ao dissociados e neutralizados ao colidem com mol\'eculas de g\'as hidrog\^enio, dando origem
a \'atomos de hidrog\^enio excitados que se movem com velocidade $v$, e tamb\'em a \'atomos excitados em repouso. 
A luz resultante da transi\c{c}\~ao $H_{\beta}$ na s\'erie de Balmer \'e observada no sentido do movimento
dos \'atomos de hidrog\^enio (a $0°$); no sentido oposto (a $180°$), com a ajuda de um espelho; e para \'atomos
em repouso. 
O comprimento de onda $\lambda''$, a $0º$, pode ser obtido pelo desenvolvimento de (\ref{eq.doppler1}) em s\'erie de Taylor:
\begin{eqnarray}
\lambda''=\left(1-\beta+\frac{1}{2}\beta^2-...\right)\lambda
\label{eq.doppler3}
\end{eqnarray}
Em que $\lambda$ \'e o comprimento de onda da radia\c{c}\~ao emitida pelos \'atomos de hidrog\^enio excitados e que est\~ao
em repouso. O comprimento de onda $\lambda'$, a $180º$, pode ser obtido pelo desenvolvimento de (\ref{eq.doppler3}) 
trocando o sinal de $v$:
\begin{eqnarray}
\lambda'=\left(1+\beta+\frac{1}{2}\beta^2+...\right)\lambda
\label{eq.doppler2}
\end{eqnarray}
Para o caso cl\'assico, o termo com $\beta^2$ pode ser desprezado:
\begin{eqnarray}
\lambda''=\left(1-\beta\right)\lambda
\end{eqnarray}
\begin{eqnarray}
\lambda'=\left(1+\beta\right)\lambda
\end{eqnarray}
Tanto no caso relativ\'istico quanto no caso cl\'assico, \'e poss\'ivel definir $\Delta\lambda_1$:
\begin{eqnarray}
\Delta\lambda_1=\frac{\lambda'-\lambda''}{2}=\beta\lambda
\end{eqnarray}
As m\'edias entre $\lambda_1$ e $\lambda_2$, para os casos relativ\'istico e cl\'assico, s\~ao, respectivamente:
\begin{eqnarray}
\bar{\lambda}=\frac{\lambda'+\lambda''}{2}=(1+\frac{1}{2}\beta^2)\lambda
\end{eqnarray}
e
\begin{eqnarray}
\bar{\lambda}=\frac{\lambda'+\lambda''}{2}=\lambda
\end{eqnarray}
A diferen\c{c}a $\Delta\lambda_2=\bar{\lambda}-\lambda$, para os casos relativ\'istico e cl\'assico, s\~ao, respectivamente:
\begin{eqnarray}
\Delta\lambda_2=\bar{\lambda}-\lambda=\frac{1}{2}\lambda\beta^2=\frac{1}{2\lambda}(\Delta\lambda_1)^2
\label{deltalambda2}
\end{eqnarray}
e
\begin{eqnarray}
\Delta\lambda_2=\bar{\lambda}-\lambda=0
\end{eqnarray}
Os resultados obtidos no experimento de Ives-Stilwell confirmaram (\ref{deltalambda2}).

\section{Massa relativ\'istica}

Outra consequ\^encia da teoria da relatividade especial \'e que a massa de um corpo depende da sua velocidade \cite{lewis}.
Esse fen\^omeno foi comprovada experimentalmente com part\'iculas $\beta$ ainda em 1909 \cite{bucherer}.
Considere um referencial $S'$ que se move com velocidade constante $v$ ao longo do eixo $x$ em rela\c{c}\~ao
a outro referencial $S$. Jo\~ao em repouso com rela\c{c}\~ao a $S$ lan\c{c}a uma esfera
de massa $m_0$ com velocidade $u_y$. Jos\'e em repouso com rela\c{c}\~ao 
a $S'$ lan\c{c}a uma esfera id\^entica com velocidade $u'_y$. Essas esferas sofrem uma colis\~ao perfeitamente
el\'astica, portanto, Jo\~ao observa uma varia\c{c}\~ao do momento linear para a esfera que ele lan\c{c}ou de
$2m_0u_y$. Se Jo\~ao observa uma massa igual a $m_0$ para a esfera lan\c{c}ada por Jos\'e, 
de (\ref{vlorentz}) obt\'em-se que Jo\~ao observa uma varia\c{c}\~ao do momento linear para a 
esfera que Jos\'e lan\c{c}ou de $2m_0u'_y=2m_0u_y/\gamma$, o que vai contra o princ\'ipio de conserva\c{c}\~ao
do momento linear. Para manter a validade desse princ\'ipio \'e necess\'ario considerar que 
Jo\~ao observa uma massa igual a $m=\gamma m_0$ para a esfera lan\c{c}ada por Jos\'e. Logo:
\begin{equation} 
m = m_0\gamma = \frac{m_0}{ \sqrt{1 - { \frac{v^2}{c^2}}}}
\label{massarelativistica}
\end{equation} 

A massa $m_0$ \'e conhecida como massa pr\'opria ou massa de repouso da part\'icula, e $m$ \'e a massa relativ\'istica.

\section{Energia cin\'etica relativ\'istica}

A energia cin\'etica de uma part\'icula em repouso \'e igual a zero. 
Uma for\c{c}a $\vec{F}=F\hat{x}$ atuando sobre uma part\'icula ao longo do eixo $x$ 
realiza um trabalho igual \`a varia\c{c}\~ao da energia cin\'etica. Dessa forma, se essa part\'icula encontra-se inicialmente em repouso, quando
sua velocidade for igual a $v$, a sua energia cin\'etica $K$ ser\'a:

\begin{eqnarray}
K & = & K - 0 = \int_{x_{inicial}}^{x_{final}}F~dx
=\int_{x_{inicial}}^{x_{final}}\frac{d}{dt}\left(mu\right)dx
=\int_{x_{inicial}}^{x_{final}}\frac{d}{dx}\left(mu\right)\frac{dx}{dt}dx
=\int_{x_{inicial}}^{x_{final}}\frac{d}{dx}\left(mu\right)dx\frac{dx}{dt}\nonumber\\
&=&\int_{0}^{v}d\left(mu\right)u = \int_{0}^{u}\left(mdu+udm\right)u=\int_{0}^{u}\left(mudu+u^{2}dm\right)
\end{eqnarray}

De (\ref{massarelativistica}):
\begin{equation} 
m^{2}c^{2}-m^{2}u^{2}=m_{0}^{2}c^{2}\Rightarrow 2mc^{2}dm-2m^{2}udu-2u^{2}mdm=0\Rightarrow c^{2}dm=mudu+u^{2}dm
\end{equation}
Portanto:
\begin{equation} 
K=\int_{0}^{u}c^{2}dm=mc^{2}-m_{0}c^{2}\;,
\end{equation} 
Logo, a energia cin\'etica relativ\'istica \'e:
\begin{equation} 
K=\left(\gamma-1\right)m_{0}c^{2}\;.
\label{enegiacineticarelativistica}
\end{equation}

 Como era de esperar-se $K=0$ para uma part\'icula em repouso ($\gamma=1$). Para uma part\'icula de baixa velocidade
 ($v<<c$) \'e preciso substituir (\ref{fatordelorentzaproximado}) em (\ref{enegiacineticarelativistica}):

 \begin{equation} 
K=\left(1 +\frac{1}{2}  \frac{v^2}{c^2}-1\right)m_{0}c^{2}=\frac{1}{2} m_{0} {v^2},
\label{enegiacineticaclassica}
\end{equation} que corresponde \`a energia cin\'etica da mec\^anica cl\'assica. Dessa forma, para baixas velocidades
($v<<c$), a velocidade cl\'assica $v$ \'e:

\begin{equation} 
v=\sqrt{\frac{2K}{m_0}}\;.
\label{vK}
\end{equation}

A seu turno, segundo a relatividade especial a express\~ao correta \'e:

\begin{equation} 
v=c\sqrt{1-\left(1+\frac{K}{m_0c^2}\right)^{-2}}\;,
\label{vKrelativistica}
\end{equation} a qual \'e obtida substituindo \ref{fatordelorentz} em \ref{enegiacineticarelativistica} e isolando $v$.

Um experimento did\'atico que comprova esses resultados foi realizado por Bertozzi e Aron \cite{bertozzi},
nesse expe- rimento el\'etrons recebiam energias cin\'eticas em MeV
por meio de um acelerador linear, e as velocidades eram calculadas atrav\'es da medi\c{c}\~ao dos tempos necess\'arios 
para que esses el\'etrons percorressem uma dist\^ancia de $8.4~m$. 
Na figura \ref{fig.compara}, esses resultados experimentais s\~ao comparados aos obtidos 
tanto pela mec\^anica cl\'assica (\ref{vK}) quanto pela mec\^anica relativ\'istica (\ref{vKrelativistica}), utilizando
$m_0=0.511~Mev/c^2$. Analisando essa figura, v\^e-se que
os resultados experimentais al\c{c}am a mec\^anica relativ\'istica \`a condi\c{c}\~ao de teoria mais exata,
mostrando que a mec\^anica cl\'assica n\~ao \'e uma teoria de todo precisa.

\begin{figure}[!h]
\begin{center}
\includegraphics[scale=1.0, bb = 50 50 400 300]{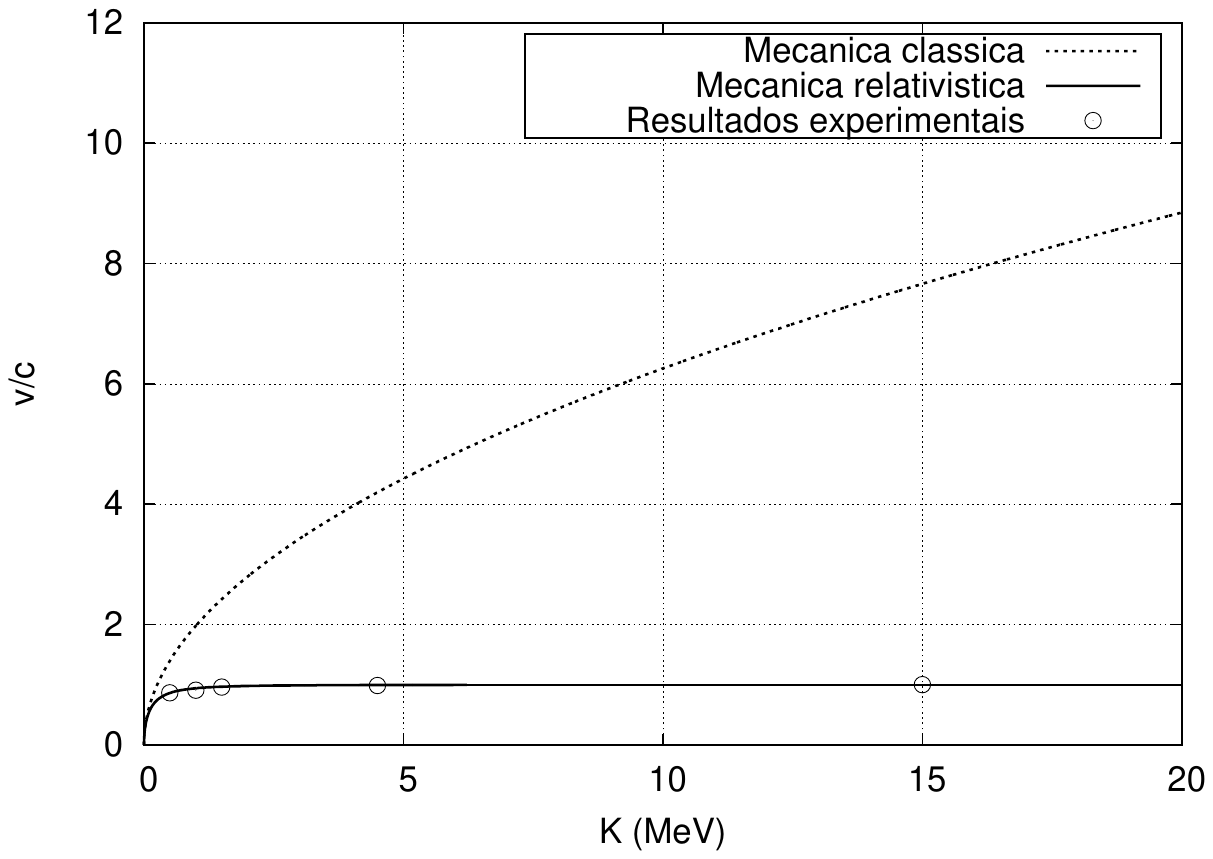}
\end{center}
\caption{Compara\c{c}\~ao de curvas K (Mev) versus v/c entre: resultados experimentais \cite{bertozzi}, previstos pela mec\^anica
cl\'assica (\ref{vK}) e previstos pela mec\^anica relativ\'istica (\ref{vKrelativistica}).}
\label{fig.compara}
\end{figure}

Fazendo o limite no infinito para (\ref{vK}):

\begin{equation} 
\lim_{K\rightarrow\infty}v=\lim_{K\rightarrow\infty}\sqrt{\frac{2K}{m_0}}=\infty\;.
\label{limvK}
\end{equation}

Por outro lado, o limite no infinito para (\ref{vKrelativistica}) \'e:

\begin{equation} 
\lim_{K\rightarrow\infty}v=\lim_{K\rightarrow\infty}c\sqrt{1-\left(1+\frac{K}{m_0c^2}\right)^{-2}}=c\;.
\label{limvKrelativistica}
\end{equation}

Comparando (\ref{limvK}) e (\ref{limvKrelativistica}), conclui-se que diferentemente do que ocorre para a mec\^anica
cl\'assica, na mec\^anica relativ\'istica existe um velocidade m\'axima (igual a $c$) que limita o movimento
das part\'iculas.

\section{Energia relativ\'istica}

E a energia relativ\'istica \'e:
\begin{equation} E=\gamma m_{0}c^{2}=mc^{2}=K+m_{0}c^{2}\;.\end{equation} 

A energia relativ\'istica tamb\'em pode ser expressa como:
\begin{eqnarray} 
 m^{2}c^{4}-m_{0}^{2}c^{4}=\gamma^{2}m_{0}^{2}c^{4}-m_{0}^{2}c^{4}=m_{0}^{2}c^{4}(\gamma^2-1) \nonumber \\
 =m_{0}^{2}c^{4}\left(\frac{1}{1-\frac{v^2}{c^2}}-1\right)
 =m_{0}^{2}c^{4}\left(\frac{\frac{v^2}{c^2}}{1-\frac{v^2}{c^2}}\right) \nonumber \\
 =\frac{m_{0}^{2}c^{2}v^2}{1-\frac{v^2}{c^2}}=m^{2}c^{2}v^2=c^2p^2 \nonumber \\
 \Rightarrow E^2=c^2p^2+m_{0}^{2}c^{4}
 \label{energiatotal}
\end{eqnarray} 

\newpage

\appendix

\section{C\'alculo de c}
\label{calculoc}

Os fen\^omenos eletromagn\'eticos s\~ao descritos pelas equa\c{c}\~oes de Maxwell:
\begin{eqnarray}
\nabla\times\vec{E}=-\frac{\partial\vec{B}}{\partial t} \label{faraday}~~~~~~~~~~\textrm{(Lei de Faraday)} \\
\nabla\times\vec{H}=\vec{J}+\frac{\partial\vec{D}}{\partial t} \label{ampere}~~~~~~~~~~\textrm{(Lei de Ampere)} \\
\nabla \cdot \vec{D}=\rho ~~~~~\textrm{(Lei de Gauss do campo el\'etrico)}\label{gausse} \\
\nabla \cdot \vec{B}=0 ~~~~~\textrm{(Lei de Gauss do campo magn\'etico)}\label{gaussh}
\end{eqnarray} onde $\vec{E}$ e $\vec{H}$ s\~ao, respectivamente, os campos el\'etrico e magn\'etico;
$\rho$ e $\vec{J}$ s\~ao, respectivamente, as densidades de carga e de corrente;
e $\vec{D}$ e $\vec{B}$ s\~ao, respectivamente, as densidades de fluxo el\'etrico e magn\'etico.
 
Utilizando a identidade vetorial $\nabla\cdot(\nabla\times\vec{H})=0$ em (\ref{ampere}), obt\'em-se:
\begin{equation}
\nabla\cdot\vec{J}=-\frac{\partial\rho}{\partial t} \label{continuidade}
~~~~~\textrm{(Eq. da continuidade)}
\end{equation}

Para um meio linear de permissividade el\'etrica relativa $\varepsilon_r$, permeabilidade
magn\'etica $\mu_r$ e condutividade el\'etrica $\sigma$, valem as rela\c{c}\~oes constitutivas:
\begin{eqnarray}
\vec{D}=\varepsilon_r\varepsilon_0\vec{E}, \label{densidadedefluxoeletrico} ~~~~~~
\vec{B}=\mu_r\mu_0\vec{H} \label{densidadedefluxomagnetico} ~~~~~~ \textrm{e} ~~~~~~
\vec{J}=\sigma\vec{E}, \label{densidadedecorrente}
\end{eqnarray} onde $\varepsilon_0 = 8,85 \times 10^{-12} F/m$ e $\mu_0 = 1,26 \times 10^{-6} A/m$.
Para o v\'acuo $(\varepsilon_r=1$, $\mu_r=1$ , $\sigma=0)$, na aus\^encia de cargas 
e correntes $\rho=0$ e $\vec{J}=0$, utilizando a identidade vetorial 
$\nabla\times(\nabla\times\vec{E})=\nabla(\nabla\cdot\vec{E})-\nabla^2\vec{E}$, e as equa\c{c}\~oes (\ref{faraday}),
(\ref{ampere}), (\ref{gausse}) e (\ref{densidadedecorrente}) , obt\'em-se:
\begin{equation}
\nabla^2\vec{E}=\frac{1}{{\mu_0\varepsilon_0}^2}\frac{\partial^2\vec{E}}{\partial t^2}=\frac{1}{c^2}\frac{\partial^2\vec{E}}{\partial t^2}
~~~~~~\textrm{(Equa\c{c}\~ao de onda)} \label{ondae}
\end{equation}
Al\'em disso, utilizando a identidade vetorial 
$\nabla\times(\nabla\times\vec{H})=\nabla(\nabla\cdot\vec{H})-\nabla^2\vec{H}$, e as equa\c{c}\~oes (\ref{faraday}),
(\ref{ampere}), (\ref{gaussh}) e (\ref{densidadedecorrente}) , obt\'em-se:
\begin{equation}
\nabla^2\vec{H}=\frac{1}{{\mu_0\varepsilon_0}^2}\frac{\partial^2\vec{H}}{\partial t^2}=\frac{1}{c^2}\frac{\partial^2\vec{H}}{\partial t^2}
~~~~~~\textrm{(Equa\c{c}\~ao de onda)} \label{ondah}
\end{equation}
Em equa\c{c}\~oes da forma (\ref{ondae}) e (\ref{ondah}), $c$ representa
a velocidade da onda, portanto, a onda eletromagn\'etica propaga no v\'acuo com velocidade $c$ igual a:
\begin{equation}
c=\frac{1}{\sqrt{\mu_0\varepsilon_0}} \label{velocidadedaluz}=299~792~458~m/s
\end{equation} 

\section{Press\~ao de radia\c{c}\~ao}
\label{calculop}

Considere uma onda plana com os campos el\'etrico (de magnitude $E$) e magn\'etico (de magnitude $B$) 
pola- rizados nas dire\c{c}\~oes $x$ e $y$, respectivamente, e que a onda desloca-se na dire\c{c}\~ao $z$, 
at\'e atingir a superf\'icie de um objeto constituido por um material altamente resistivo.
O campo el\'etrico move o el\'etron de carga $-e$ com uma for\c{c}a $F_E=-eE$. A alta resistividade resulta em um
fator de amortecimento consider\'avel, capaz de fazer com que uma for\c{c}a dissipativa leve o movimento ao equil\'ibrio, 
fazendo o el\'etron mover-se com velocidade constante $v_e$. Assim, esse el\'etron estar\'a se movimentando numa dire\c{c}\~ao
perpendicular ao campo magn\'etico de intensidade $B$, e portanto, uma for\c{c}a magn\'etica $F_z=-ev_eB$ estar\'a sendo exercida
sobre o objeto. 

Se $U_e$ \'e a energia absorvida pelo el\'etron, ent\~ao:

\begin{equation} 
\frac{dU_e}{dt} = F_Ev_e=-eEv_e
\label{potenciae}
\end{equation} 

Por sua vez, se $p_e$ \'e o momento linear exercido sobre o el\'etron e dado que $B=E/c$, tem-se:

\begin{equation} 
\frac{dp_e}{dt} = -ev_eB=-ev_e\frac{E}{c}=\frac{1}{c}\frac{dU_e}{dt}
\end{equation} 

\begin{equation} 
\int_{0}^{t}\frac{dp_e}{dt}dt = \frac{1}{c}\int_{0}^{t}\frac{dU_e}{dt}dt
\end{equation} 

\begin{equation} 
p_e = \frac{U_e}{c}
\label{forcae}
\end{equation}

}
\Teil{Bibliography}{
\addcontentsline{toc}{section}{\refname}
\providecommand{\href}[2]{#2}\begingroup\raggedright\endgroup

\fbox{
\begin{tabular}{l}
  Tiago Carvalho Martins is currently professor at the \textit{Universidade Federal do Sul e Sudeste do Par\'a}, \\
  Marab\'a, Brazil. Dr. Tiago Carvalho Martins is interested in computational physics and optimization\\
  techniques applied to electromagnetic problems. \\
\end{tabular}
}

}
\end{document}